\documentclass[%
 reprint,
nofootinbib,
 aps
]{revtex4-2}

\usepackage[utf8]{inputenc}
\usepackage{bbm}
\usepackage{appendix}

\usepackage[usenames,dvipsnames]{xcolor}
\usepackage{color}
\usepackage{colortbl}%

\definecolor{qctrl_primary}{HTML}{680Ce9}
\definecolor{qctrl_secondary}{HTML}{BF04DC}
\definecolor{qctrl_noise}{HTML}{7B7479}
\definecolor{qctrl_axis_labels}{HTML}{514B4F}
\definecolor{qctrl_borders}{HTML}{CFCBCE}
\definecolor{qctrl_blue}{HTML}{4177D8}
\definecolor{qctrl_aqua}{HTML}{32A4A8}
\definecolor{qctrl_green}{HTML}{32A857}
\definecolor{qctrl_lime_green}{HTML}{A2A933}
\definecolor{qctrl_orange}{HTML}{D6742F}
\definecolor{qctrl_red}{HTML}{D84144}
\definecolor{qctrl_fuchsia}{HTML}{D84190}

\usepackage{blindtext}

\usepackage{amssymb,amsthm}
\usepackage{amsmath}
\usepackage{mathtools} %

\usepackage{mathrsfs}
\usepackage{dsfont} 
\usepackage[nice]{nicefrac}

\usepackage{cancel}
\usepackage{bm}%

\RequirePackage{algorithm}
\RequirePackage{algpseudocode}%

\theoremstyle{definition}

\theoremstyle{remark}

\usepackage{physics}
\usepackage{siunitx} %

\usepackage{qcircuit}
\usepackage{braket}
\usepackage{graphicx}%
\usepackage{float}
\usepackage{rotating}
\usepackage{subcaption}  %
\usepackage{caption}
\usepackage{tikz}        %
\usepackage{svg}
\usepackage{enumitem}
\setlist{nosep} 

\usepackage{dcolumn}%
\usepackage{multirow}
\usepackage{makecell}
\usepackage{bigdelim}
\usepackage{blkarray}
\usepackage{array}

\usepackage{hyperref}
\hypersetup{
final=true,
plainpages=false,
pdfstartview=FitV,
pdftoolbar=true,
pdfmenubar=true,
bookmarksopen=true,
bookmarksnumbered=true,
breaklinks=true,
linktocpage,
colorlinks=true,
linkcolor=blue,
filecolor=blue,      
urlcolor=cyan,
citecolor=blue,
anchorcolor=green
}

\usepackage[nameinlink]{cleveref}
\crefname{equation}{Eq.}{Eqs.}
\crefname{align}{Eq.}{Eqs.}
\crefname{figure}{Fig.}{Figs.}
\crefname{table}{Table}{Tables}
\crefname{tabular}{Table}{Tables}
\crefname{section}{Sec.}{Secs.}
\crefname{appendix}{App.}{Apps.}

\crefname{appsec}{App.}{Apps.}
\crefname{appchapter}{App.}{Apps.}

\crefname{algorithm}{Alg.}{Algs.}
\creflabelformat{equation}{#2#1#3}

\usepackage{natbib}
\DeclareMathOperator{\can}{\mathcal{C}}

\usepackage{soul}
\usepackage{layouts}
\usepackage{diagbox}
\begin{document}
\title{No need to calibrate: characterization and compilation for high-fidelity circuit execution using imperfect gates}

 \author{Ashish Kakkar}
 \altaffiliation{These authors contributed equally}
 \email{samuel.marsh@q-ctrl.com}
 \author{Samuel Marsh} 
 \altaffiliation{These authors contributed equally}
 \email{samuel.marsh@q-ctrl.com}
 \author{Yulun Wang}
 \author{Pranav Mundada}
 \author{Paul Coote}
 \author{Gavin Hartnett}
 \author{Michael J. Biercuk}
 \author{Yuval Baum}
\affiliation{Q-CTRL, Los Angeles, CA USA and Sydney, NSW Australia}
\date{\today}

\begin{abstract}
We propose and validate on real quantum computing hardware a new method for extended two-qubit gate set design, replacing iterative, fine calibration with fast characterization of a small number of gate parameters which are then tracked and corrected in circuit compilation.
Coherent contributions to the pulse unitary that would traditionally be considered sources of error are treated as part of the gate definition, and compensated in software via single-qubit rotations.
This approach enables rapid device-wide generation of high-fidelity two-qubit entangling gates, which are combined with standard calibrated gates to produce an expanded gate set.
We show how these gates are directly usable as part of a quantum compiler, synthesizing generic two-qubit circuit blocks into minimal-duration sequences of the characterized gates interleaved with compensating single-qubit rotations.
Benchmarking against circuits compiled using the default $CX$ gate alone on 127-qubit IBM hardware shows up to 7X improvement in success probability for Quantum Fourier Transform circuits up to 26 qubits, and up to 9X lower mean-square error in Trotter simulations of the one-dimensional transverse-field Ising model.
Our hardware-agnostic characterization and compilation methodology makes it practical to scale up expressive gate sets on quantum computing architectures while minimizing the need for onerous fine-tuning of low-level control waveforms.
\end{abstract}

\maketitle
\section{Introduction}

Quantum information processing relies on a bidirectional relationship between logical gate operations and low-level control pulses which mediate physical interactions between qubits and control fields: compilation maps gates down to analog pulses, while gate set design turns pulse sequences into logical operations.
Because universal quantum computation hinges on the accurate synthesis of two-qubit unitaries~\cite{Kitaev1997, DiVincenzo1995, Lloyd1995}, the performance of two-qubit gates is a central bottleneck.
A two-qubit gate that generates any entanglement at all is sufficient for universality, but in practice particular classes of gates are more efficient than others.
Special perfect entangling gates, such as the $CX$ gate, are an important class of gates that can synthesize any two-qubit operation using at most three applications~\cite{Zhang2005}.
There has been remarkable progress in designing fast entangling two-qubit gates, suppressing parasitic coherent errors using both pulse-level and circuit-level techniques, and using a wide range of entanglement-generating interactions~\cite{Majer2007, DiCarlo2009, Rigetti2010, Chow2013, Sheldon2016, Casparis2019,  Wei2022,Lin2022,Lao2022}.

Extending an existing universal gate set with additional entangling gates which have lower entangling power but shorter duration can reduce overall algorithm duration and boost fidelity in practice~\cite{Abrams2020,Warren2023}.
In particular, parametric gates~\cite{Abrams2020, Foxen2020, Hill2021, Collodo2020} are a common addition to quantum computing gate sets, typically relating the logical gate characteristics  to the physical control parameters (e.g. pulse duration) through an experimental calibration process.
This enables shorter-duration execution via choosing the minimal gate parameters required to synthesize a given unitary~\cite{Stenger2021,Earnest2021,Sugawara2025}. An alternative approach involves extending the two-qubit gate set via calibration of an additional number of discrete entangling gates.
\citet{Peterson2022} show through numerical simulations that augmenting the standard $CX$ gate with only two shorter-duration calibrated gates, $CX^{1/2}$ and $CX^{1/3}$, yields random two-qubit unitary synthesis performance within 15\% of using the full continuous family $CX^\alpha$ for all $\alpha \in [0,1]$.

However, both approaches to constructing such gates pose significant challenges.
First, calibrating a two-qubit interaction to a target unitary with high fidelity is resource-intensive~\cite{Baum2021, Ball2021, Machnes2018, Klimov2020, Khaneja2005}, and this challenge is exacerbated when a nonlinear relationship between logical gate characteristics and physical pulse parameters must be determined. Next, hardware providers face a demanding tradeoff between the positive impact of additional calibration efforts and associated penalties on device uptime, especially those without pulse-level APIs to enable users/third parties to customize gate waveforms or calibrate additional gate definitions~\cite{Baum2021,Carvalho2021}.

In this work, we introduce a new approach for extending an entangling gate set without the need for additional calibration or extensive analog-layer programming.
Our approach to diversifying the entangling gate set is based on a shift of perspective: rather than \textit{calibration}, we focus on \textit{characterization and compilation}.
That is, instead of calibrating a specific target gate by optimizing a pulse sequence or waveform,
we construct a pulse sequence to approximately realize a desirable gate (such as $CX^{1/3}$) but do not attempt to refine the pulse definition via hardware-in-the-loop calibration.
Rather, we perform a narrow set of experiments to precisely learn key aspects of the associated unitary, leveraging tomographic techniques~\cite{Nielsen2021,Sheldon2016, Sundaresan2020, Wei2024, Gross2024}, and then (irrespective of what the unitary is in practice) we account for this definition in software. We introduce an end-to-end pipeline that efficiently characterizes arbitrary pulse sequences to create additional entangling gates, and leverages this extended gate set to compile circuits with improved duration and fidelity. Our method determines the shortest-duration sequence of characterized pulses that can implement the target unitary. In real demonstrations on IBM hardware, we observe up to 7X higher success probability for the quantum Fourier transform for up to 26 qubits compared to standard $CX$ synthesis, and up to 9X lower mean-square error (MSE) in expectation value estimation for a Trotterized Hamiltonian simulation of the one-dimensional transverse field Ising model on 25 qubits.

The manuscript is organized as follows.
We give a brief introduction to the necessary background information in \cref{sec:background}.
Then \cref{sec: Compiler architecture design for efficient gates} describes how efficient gate tomography techniques can be used to produce characterized pulses, and how these arbitrary pulses can be utilized by a compiler to efficiently synthesize two-qubit unitaries appearing in quantum circuits.
We then consider special cases for characterization and compilation in \cref{sec:special-cases}, where both steps can be accelerated compared to the general case, and where synthesis conditions and relevant circuit parameters can be determined in closed form.
Finally, in \cref{sec: Experimental Algorithmic Benchmarking}, we demonstrate the benefits of this fully automated technique through algorithmic benchmarking on IBM's superconducting devices.

\section{Background on two-qubit gates and the Weyl chamber}
\label{sec:background}

Any unitary operation on two qubits can be decomposed into a product of local operations and a canonical entangling interaction using the Cartan decomposition, which expresses a general $SU(4)$ matrix as
\begin{equation}
    U = (w_1 w_2) \cdot \can(c_1, c_2, c_3) \cdot (w_3 w_4) \, .
\end{equation}
Here, each $w_i \in SU(2)$ is a local (single-qubit) operation, and
\begin{equation}
    \label{can notation}
    \can(c_1, c_2, c_3) = \exp\left( -\frac{i}{2}(c_1 XX + c_2 YY + c_3 ZZ)\right)
\end{equation}
is the canonical gate. The parameters $\mathbf{c}=(c_1, c_2, c_3)$ are referred to as the canonical coordinates; these describe the fundamental nonlocal content of the gate. Through permutation and reflection symmetries, the set of all such equivalence classes can be restricted to a specific tetrahedral region in $\mathbb{R}^3$ known as the \textit{Weyl chamber}. This region is described by the inequalities $c_1 \geq c_2 \geq c_3 \geq 0$ and $c_1 + c_2 \leq \pi$. 

The Weyl chamber therefore gives a geometric representation of all two-qubit gates up to local equivalence. Two gates whose Cartan decomposition leads to the same point in the Weyl chamber are said to be \textit{locally equivalent}, meaning they differ only up to single-qubit unitaries. Each point within the chamber represents an equivalence class of gates all locally equivalent to each other. We write $U\sim V$ to denote local equivalence between $U$ and $V$. 

An important category of gates within the Weyl chamber is the set of perfect entanglers, which are gates capable of generating a maximally entangled state from a product state. A subset of these gates, known as special perfect entanglers (such as the $CX$ gate), take an orthonormal product basis to an orthonormal maximally entangled basis. Special perfect entanglers can be used to synthesize any two-qubit unitary using at most three applications (and in the case of the $B$ gate, at most two applications~\cite{Zhang2004}) when interleaved with single-qubit operations~\cite{Rezakhani2004,Zhang2005}.

The Weyl chamber and associated geometric relationships provide a natural framework for the synthesis of two-qubit gates. Rather than directly aiming to reproduce a target $SU(4)$ unitary, a compiler can aim to synthesize the associated point $\mathbf{c}=(c_1, c_2, c_3)$ (i.e. any unitary with the same canonical coordinates), and then use the Cartan decomposition to determine the appropriate single-qubit unitaries to transform the synthesized point precisely into the target unitary. This concept forms the basis for the characterization and compilation methodology that we explore within this work. For more background information on the Weyl chamber and the Cartan decomposition, refer to Refs.~\cite{Zhang2003,Tucci2005}.

Having introduced the representation of two-qubit gates in the Weyl chamber, we now demonstrate how this framework directly informs our proposed pulse-aware compiler architecture.

\section{Pulse characterization and pulse-aware compiler design}\label{sec: Compiler architecture design for efficient gates}

\begin{figure*}[!t]
    \centering    \includegraphics[width=\linewidth]{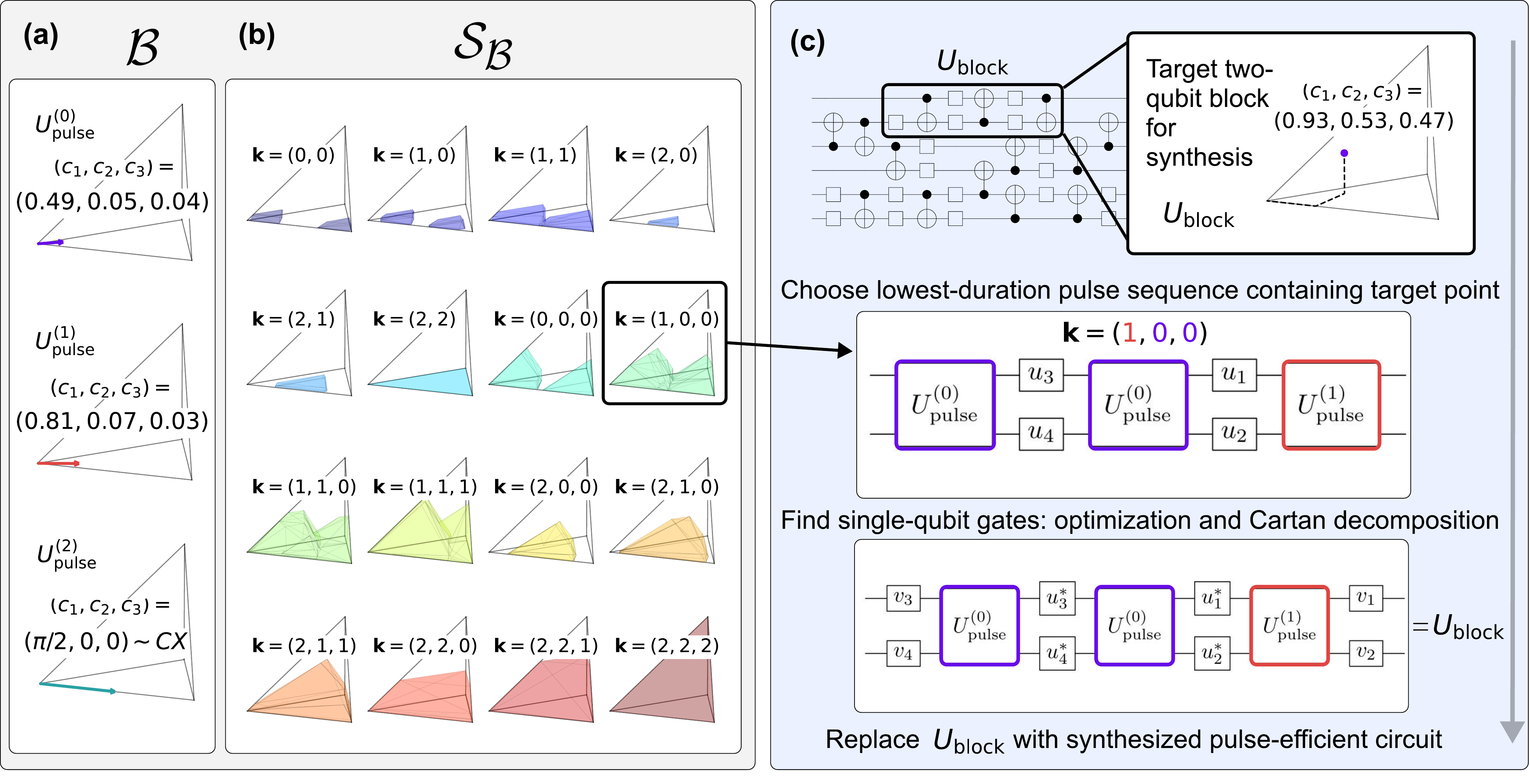}
    \caption{
    The pulse-efficient compilation procedure. \textbf{(a)} An extended two-qubit gate set $\mathcal{B}$ for a specific pair of qubits. Each gate is represented by a point in the Weyl chamber (the space of all two-qubit unitaries up to local equivalence). Here, $\mathcal{B}$ contains a calibrated $CX$ gate $U_\text{pulse}^{(2)}$, along with two shorter gates that were characterized as $U_\text{pulse}^{(0)}$ and $U_\text{pulse}^{(1)}$. \textbf{(b)} The corresponding coverage set $\mathcal{S}_\mathcal{B}$, representing the regions of the Weyl chamber reachable by different combinations of these gates interleaved with single-qubit gates. We organize the coverage set by the number of applications of each two-qubit entangling gate, encoded as a vector $\bm{k} = (k_{0}, \ldots, k_{n-1})$ having entries ordered right-to-left with respect to circuit ordering, where $k_i \in \{0,1,2\}$ indicates which of the three entangling gates is used. For example, $\mathbf{k}=(1, 0, 0)$ denotes the space of unitaries reachable by two applications of $U_\text{pulse}^{(0)}$ followed by $U_\text{pulse}^{(1)}$, interleaved with single-qubit gates. Although three applications of $CX$ ($\mathbf{k}=(2, 2, 2)$) is universal, many unitaries can be synthesized with other lower-duration combinations of the gate set. \textbf{(c)} During compilation, when presented with a two-qubit block, the lowest-duration pulse sequence containing the coordinates of that block is chosen. Inner single-qubit gates are computed via numerical optimization and outer gates are computed by the Cartan decomposition, producing a synthesis of $U_\text{block}$ using gates drawn from $\mathcal{B}$.}
    \label{fig:synthesis-overview}
\end{figure*}

In this section, we present our pulse-efficient characterization and compilation methodology in generality. Our approach is summarized in \cref{fig:synthesis-overview}, illustrating an example of how to use two additional efficient entangling pulses alongside the calibrated $CX$ gate to compile circuits into lower-duration equivalents.

\subsection{Obtaining an extended gate set through characterization}

Our motivation for characterization is to generate an extended two-qubit entangling gate set $\mathcal{B} = \{U_\text{pulse}^{(i)}\}$ of $SU(4)$ matrices for each connected qubit pair, composed of the calibrated special perfect entangler from the original gate set along with a desired number of \textit{characterized pulses}. The extended gate set $\mathcal{B}$ will be different for every connected pair of physical qubits on the hardware, something easily tracked in software.  We define a characterized pulse (which we refer to interchangeably as a ``pulse'' throughout this work) as any waveform/control on a pair of qubits that induces an entangling unitary operation, and where we have performed some tomographic procedure to characterize the associated unitary matrix.

To generate these additional pulses, we use the control sequence that generates the original calibrated entangling gate, rescaling it to produce one or more shorter-duration waveforms.
These pulses will have lower entangling power than the original special perfect entangler, but can be more efficient to utilize in a quantum circuit.

To this end, for a given interaction modality on a given quantum computing architecture, we assume there is a method of scaling the effective rotation angle of the calibrated gate.
Typically, this consists of reducing or transforming the durations and/or amplitudes of the pulse shape(s) that comprise the calibrated gate.
The specific pulse-level heuristics depend on the relevant device physics, but often involve linear scaling of pulse durations to proportionally reduce the integral of the pulse envelope.
In general, we expect some method of pulse scaling to be available for the majority of interaction modalities, enabling shorter-duration pulses for which a unitary operation is obtained.

Furthermore, it is typical that we can approximate the entangling power of a rescaled waveform a priori, using a theoretical model of the physics behind the interaction.
This enables \textit{approximate} targeting of a particular point in the Weyl chamber (e.g. corresponding to a desirable gate such as $CX^{1/2}$ or $CX^{1/3}$).
We provide one such example of this rescaling idea applied to the cross-resonance pulse for superconducting qubits in~\cref{sec:special-cases}.

Given a rescaled waveform, the next step is to precisely characterize the unitary $U_\text{pulse}$ corresponding to this waveform.
We can leverage the assumption that the pulse is unitary in the two-qubit computational subspace of the physical system, and use well-known approaches for process tomography under the unitarity assumption.
With this assumption, process tomography of unitary maps enables a near-quadratic reduction in the number of required measurement settings in the Hilbert space dimension~\cite{Shabani2011,Flammia2012,Gutoski2014,Baldwin2014}. In two-qubit experiments, unitary/near-unitary gates have been reconstructed to high fidelity with only tens of configurations instead of the hundreds required by standard process tomography~\cite{Shabani2011}.

Additionally, typically the known theoretical model underlying a two-qubit interaction will further restrict the degrees of freedom available in the unitary, which can be leveraged for faster tomography. We provide a specialized characterization process for a restricted class of two-qubit unitaries in~\cref{sec:special-cases}, which reduces the two-qubit pulse characterization to two simultaneous instances of single-qubit unitary tomography along with estimation of the phase difference between these single-qubit unitaries.

Thus, overall, the first key step of our pulse-efficient methodology is to rescale a calibrated gate on each qubit pair to generate one or more shorter-duration waveforms, which are then characterized and added to that qubit pair's gate set $\mathcal{B}$. With this extended gate set at hand, irrespective of what the unitaries associated with the characterized pulses turned out to be, the pulses can be utilized to \textit{exactly} compile an input circuit into a shorter-duration equivalent.
We present a compilation procedure to achieve this in the following section.

\subsection{Circuit compilation with efficient pulses}

The compilation process of converting the quantum circuit to use this extended gate set, which we call \textit{pulse-efficient compilation}, is applicable after qubit allocation and routing to account for device connectivity. The input circuit consists of one- and two-qubit gates acting on physical qubits, with all two-qubit gates acting on qubits that are physically connected in the hardware architecture.

We first identify maximal contiguous sequences of gates acting on qubit pairs. The sequences are aggregated into two-qubit \textit{blocks}, each of which is represented by an $SU(4)$ matrix, resulting in a quantum circuit represented fully by a sequence of two-qubit unitaries acting on different pairs of qubits. An illustration of a particular two-qubit block to be synthesized into a pulse-efficient gate set is shown in~\cref{fig:synthesis-overview} (c).

The core task of pulse-efficient compilation is then to re-synthesize each block into characterized pulses from $\mathcal{B}$ interleaved with single-qubit operations. Here, we describe a numerical optimization-based method for pulse-efficient compilation in full generality, supporting arbitrary pulses taken from $\mathcal{B}$.

Consider a particular two-qubit block $U_\text{block}$ in a quantum circuit. We wish to identify the optimal synthesis of this block using gates from the available set $\mathcal{B}$, along with interleaved single-qubit gates. Notably, our assumption that $\mathcal{B}$ contains at least one calibrated special perfect entangling gate (e.g. $CX$) from the original gate set ensures that any two-qubit block in a quantum circuit can be synthesized exactly using, in the worst case, at most three gates from $\mathcal{B}$.

\begin{figure*}[!ht]
    \centering
    \includegraphics[width=\linewidth]{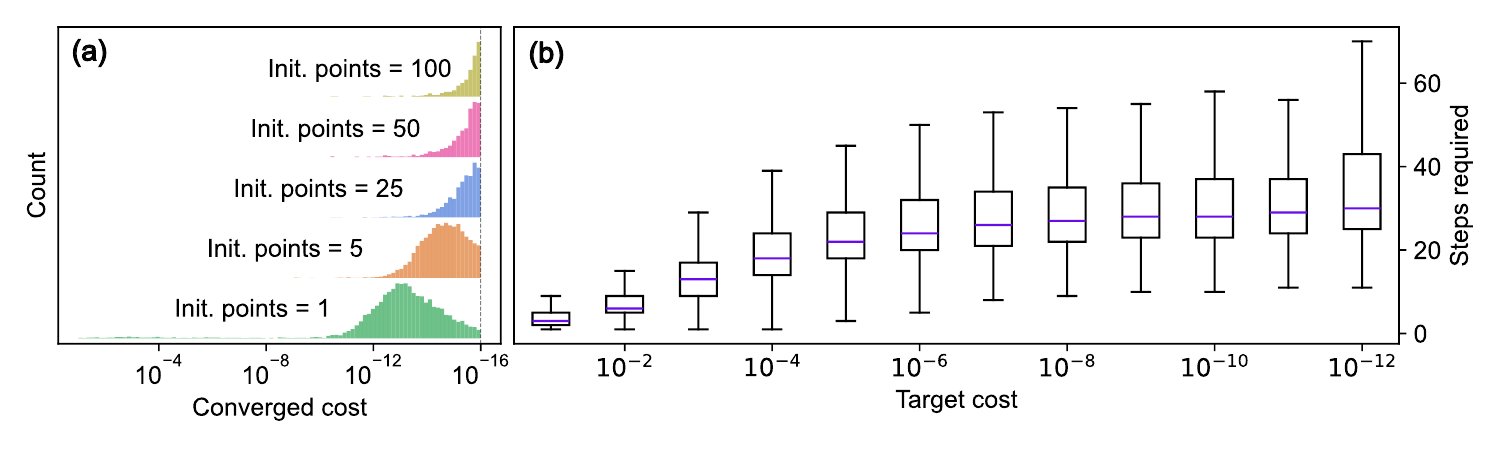}
    \caption{The robustness and efficiency of the numerical invariant-matching synthesis scheme for randomly generated targets, using the Q-CTRL Boulder Opal optimizer~\cite{Ball2021}. \textbf{(a)} Distributions of the minimum converged cost for different numbers of initial points, from 10,000 random instances. \textbf{(b)} Box plot of the total number of optimization steps required to reach different target thresholds using a retry-if-fail approach, over 10,000 random instances.}
    \label{fig:numerical-experiments-invariant-matching}
\end{figure*}

We frame optimality as minimizing the total duration of the block's implementation on hardware. Thus, with each $U_\text{pulse}^{(i)} \in \mathcal{B}$, we associate a cost which is the duration of the pulse plus the duration of performing an arbitrary $SU(2)$ rotation on both qubits. This cost metric ensures that the duration of interleaved single-qubit layers is also taken into account.

The compilation strategy follows a three-stage approach. First, we identify an appropriate ansatz, involving the selection of an optimal (lowest-cost) sequence of 1-3 gates from $\mathcal{B}$ for which it is possible to synthesize $U_\text{block}$. Second, we figure out the \textit{inner} single-qubit gates required to make this ansatz locally equivalent to $U_\text{block}$. Third, we identify the necessary \textit{outer} single-qubit gates to complete the synthesis of $U_\text{block}$.

Specifically, the ans\"atze we consider in the first compilation step are of the form
\begin{equation}
    \label{eq:synthesis-ansatz-general}
    V_{\mathbf{k}}(u) = \begin{cases}
        U_\text{pulse}^{(k_1)} & |\mathbf{k}|=1 \, ,\\
        U_\text{pulse}^{(k_1)} \cdot \left(u_1  u_2 \right) \cdot U_\text{pulse}^{(k_2)} & |\mathbf{k}|=2 \, ,\\
        U_\text{pulse}^{(k_1)} \cdot \left(u_1  u_2 \right) \cdot U_\text{pulse}^{(k_2)} \cdot \left(u_3  u_4 \right) \cdot U_\text{pulse}^{(k_3)} & |\mathbf{k}|=3 \, .
    \end{cases}
\end{equation}
Here, the tuples $\mathbf{k}$ represent choices of two-qubit gates from $\mathcal{B}$. For example, if we have $\mathcal{B} = \{U, CX\}$ for some characterized pulse $U$, then $\mathbf{k}=(2, 2, 1)$ would represent an ansatz which uses an application of $U$ followed by two $CX$ gates  (interleaved with single-qubit gates). Here, the ordering of the two-qubit gates in the ansatz does not affect the synthesis power, i.e. in this example $\mathbf{k}=(1, 2, 1)$ or any other permutation would be equally powerful~\cite{McKinney2025}.
The goal is then to find the lowest-cost $\mathbf{k}$ for which there exists single-qubit gates $u$ such that $V_\mathbf{k}(u) \sim U_\text{block}$.

We can determine whether it is possible to synthesize a given $U_\text{block}$ using $V_\textbf{k}(u)$ via the techniques established by~\citet{Peterson2020}. Associated with each $\mathbf{k}$-tuple is a geometric \textit{polytope} (generated by a set of constraints on the canonical coordinates) in the three-dimensional local invariant space, representing the volume that the $V_\textbf{k}(u)$ ansatz can synthesize for different choices of local unitaries $u$. Verifying if $V_\textbf{k}(u)$ can synthesize $U_\text{block}$ is a matter of checking if the associated polytope contains the coordinate corresponding to $U_\text{block}$.

The procedure in \cite{Peterson2020} allows us to determine the best ansatz efficiently.
Immediately after the gate set $\mathcal{B}$ is initially constructed from pulse characterization, we can determine a \textit{coverage set} $\mathcal{S}_\mathcal{B}$ and store it for use in the compilation process.
The coverage set is the collection of synthesis ans\"atze and their corresponding polytopes, $\mathcal{S}_\mathcal{B} = \{(\bm{k}, q_{\bm{k}}) : \bm{k} \in \{0,1,2\}^n, k_i \geq k_{i+1}\},$ where $\bm{k} = (k_{0}, \ldots, k_{n-1})$ encodes the two-qubit pulse sequence and $q_{\bm{k}}$ is the corresponding polytope.
It can be computed efficiently for fixed maximum two-qubit depth instantiations, which is the case for our $\abs{\mathbf{k}} \leq 3$ restriction, and is both \textit{complete} (can synthesize every $U_\text{block} \in SU(4)$) and \textit{irredundant} (no redundant ans\"atze are included in the set). An example of a coverage set is shown in \cref{fig:synthesis-overview}~(b).

Using the coverage set makes identifying the appropriate ansatz to use extremely efficient and straightforward.
Given a block $U_\text{block}$, identify the associated canonical coordinates $\mathbf{c}$. Then, iterate through the coverage set in order of ascending cost, and choose the ansatz $V_\text{k}(u)$ where the corresponding polytope contains the point $\mathbf{c}$.

However, despite the polytope representation providing an elegant mechanism to choose the optimal $\mathbf{k}$ (and thus $V_\textbf{k}(u)$) for a given block, it is non-constructive in terms of actually obtaining the appropriate single-qubit $u$ to realize a desired point within the polytope.
Thus, for the lowest-cost pulse selection $\mathbf{k}$ that can synthesize $U_\text{block}$, the second compilation step is to optimize $u$ to match the corresponding \textit{Makhlin invariants} via numerical optimization.

The Makhlin invariants are two invariants ${\mathbf{g}(U)=(g_1(U), g_2(U)) \in \mathbb{C} \times \mathbb{R}}$ that characterize local equivalence, with their explicit definition given in \cref{app:makhlin-invariants}.
The Makhlin invariants are preferred over the canonical coordinates for our purposes in this compilation step because they avoid discontinuities and branch cuts. We optimize the $u_i \in SU(2)$ to minimize the difference in invariants,
\begin{equation}
    u^* = \arg\min_{{u}} \norm{\mathbf{g}(V_{\mathbf{k}}({u})) - \mathbf{g}(U_\text{block})}^2 \, .
    \label{eq:makhlin-invariant-loss}
\end{equation}
Note the solution $u^*$ is not unique in general, with many such possible $u^*$ producing the same canonical coordinates. The landscape is non-convex, but as we will show, easily navigable by an optimizer.  If the optimization is successful to within a desired cost threshold, then an appropriate set of single-qubit gates $u^*$ has been found such that $V_\mathbf{k}(u^*) \sim U_\text{block}$. 

Given $u^*$, the third and final step is straightforward, where the required $v \in SU(2)$ such that $U_\text{block} = (v_1  v_2) \cdot V_\mathbf{k}(u^*) \cdot (v_3  v_4)$ can be computed directly via the Cartan decomposition. Applying the Cartan decomposition to both the ansatz $V_\textbf{k}(u^*)$ and the block,
\begin{align}
    U_\text{block} &= (w_1  w_2) \cdot \can(\mathbf{c}) \cdot (w_3  w_4) \, , \\
    V_\mathbf{k}(u^*) &= (w_1'  w_2') \cdot \can(\mathbf{c}) \cdot (w_3'  w_4') \, ,
\end{align}
and thus the appropriate outer unitaries to make $V_\mathbf{k}(u^*)$ equal to $U_\text{block}$ are $v_{1,2}=w_{1,2} (w_{1,2}')^{-1}$ and $v_{3,4}= (w_{3,4}')^{-1} w_{3,4}$.
Hence, the main challenge in synthesizing $U_\text{block}$ reduces to solving for local equivalence, i.e. solving for the single-qubit unitaries $u^*$ such that both $U_\text{block}$ and $V_\textbf{k}(u^*)$ share the same canonical coordinates $\mathbf{c}$.

After successful two-qubit gate synthesis, we perform conventional $SU(2)$ synthesis on the single-qubit unitaries, mapping each unitary into the native single-qubit gate set using standard techniques that do not require analog-layer pulse definitions. Commutation checking is performed to commute single-qubit gates through the characterized pulse and to the outside of the block where possible, to enable merging and optimization of single-qubit operations between different blocks. We summarize the overall two-qubit compilation protocol in \cref{alg:synthesis}.

\begin{figure}[ht!] 
\begin{algorithm}[H]
\caption{Two-qubit block compilation} 
\begin{algorithmic}[1]
 \Procedure{Synthesize}{$U_\text{block}, \mathcal{S}_\mathcal{B}$}
    \State $p \gets$ invariant coordinates of $U_\text{block}$
    \State sort coverage set $\mathcal{S}_\mathcal{B}$ by cost of $\mathbf{k}$
    \ForAll{$(\mathbf{k}, q_\mathbf{k}) \in \mathcal{S}_\mathcal{B}$}
        \If{polytope $q_\mathbf{k}$ contains point $p$}
            \State $V_\mathbf{k}(u) \gets$ define block ansatz
            \State $u^* \gets \arg\min_\mathbf{u} \norm{\mathbf{g}(V_{\mathbf{k}}({u})) - \mathbf{g}(U_\text{block})}^2$
            \State $v \gets$ Cartan decompose $U_\text{block}$ and $V_{\mathbf{k}}(u^*))$
            \State \Return $U_\text{block} \equiv (v_1  v_2) \cdot V_{\mathbf{k}}({u}^*) \cdot (v_3  v_4)$
        \EndIf
    \EndFor
  \EndProcedure
 \end{algorithmic} 
 \label{alg:synthesis}
 \end{algorithm}
\end{figure}

\begin{figure*}[!ht]
    \centering
    \includegraphics[width=\linewidth]{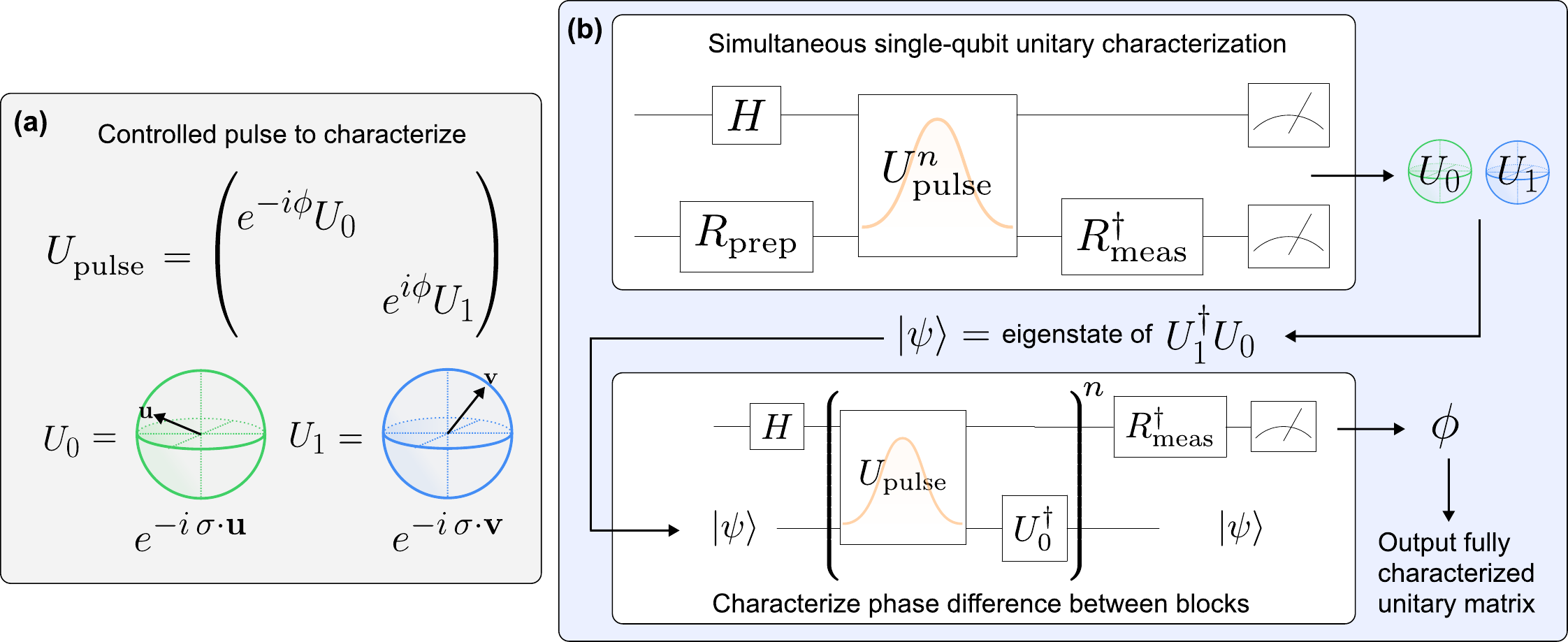}
    \caption{A protocol for complete, robust and efficient characterization of a controlled pulse unitary $U_\text{pulse}$. \textbf{(a)} A controlled pulse unitary consists of two $SU(2)$ matrices on the block diagonal, separated by an unknown phase. \textbf{(b)} The process for characterizing a target $U_\text{pulse}$. Single-qubit gate tomography, via any appropriate preparation and measurement sets, e.g. $R_\text{prep}, R_\text{meas} \in \{I, H, S.H\}$, is used to characterize the blocks $U_0$ and $U_1$ ``simultaneously'', i.e. using the same experimental configuration for each. This enables the construction of the state $\ket{\psi}$, which is applied to estimate the phase difference $2\phi$ between the two blocks via measurement of the first qubit in the $X$ and $Y$ bases ($R_\text{meas} \in \{H, S.H\}$). Curve fitting the sinusoidal relationships of different pulse iteration counts $n$ against the unitary parameters enables more accurate estimation via robustness to noise and amplification of small parameters.}
    \label{fig:characterization}
\end{figure*}

Here we validate this generic pulse-based compilation method via numerical experiments, using the gradient-based optimizer in Q-CTRL's Boulder Opal~\cite{Ball2021}. Specifically, we generate 10,000 random target unitaries by composing three Haar-random $SU(4)$ unitaries with two random interleaved $SU(2) \otimes SU(2)$ unitaries.
Taking these same three $SU(4)$ unitaries as our pulse sequence, we optimize over $u_i = Z(\alpha_i)Y(\beta_i)Z(\gamma_i) \in SU(2)$ with the aim of recovering the invariants of the original unitary.
Using a cost threshold of $10^{-12}$, a maximum of $100$ steps per optimization, and $5$ initial points, invariant matching is successful to the desired threshold in 99.83\% of cases. In practice on current quantum hardware, it would be sufficient to choose far higher cost threshold (e.g. even $10^{-5}$ could be considered highly conservative based on error rates of current state-of-the-art single-qubit gates). However, we illustrate results on extremely low cost thresholds to demonstrate the high synthesis efficiency even to numerical precision limits. In \cref{fig:numerical-experiments-invariant-matching}, we provide a detailed visualization of the convergence performance for different numbers of initial points and for different target costs. \cref{fig:numerical-experiments-invariant-matching} (b) specifically shows that the convergence is efficient, requiring only a modest number of computational steps which grows slowly from $\sim25$ to $30$ over seven decades of target cost. These results show that the optimization-based method reliably, and in very few steps, synthesizes a block from arbitrary pulses, when a solution exists.

\section{Special cases: accelerating characterization and compilation}
\label{sec:special-cases}

In this section, we explore situations where both the characterization stage
and the compilation stage
can be further enhanced in speed and robustness compared to the general case.
We first present a characterization technique for ``controlled pulses", which act as a block-diagonal unitary up to a known change of basis, enabling robust characterization using only a low number of unique experiments and shots per experiment.
We then present a compilation technique for ``single-axis pulses'', which are pulses locally equivalent to $\mathcal{C}(c_1, 0, 0)$, where the compilation can be found in closed-form.

\subsection{Accelerated characterization: controlled pulses}
\label{sec:controlled-pulse-characterization}

On real devices, theoretical models of two-qubit interactions typically constrain the dynamics to a lower-dimensional, parameterized subset of all possible two-qubit unitaries.
In these cases, rather than using the general unitary tomography techniques discussed in \cref{sec: Compiler architecture design for efficient gates}, we can design specialized characterization techniques for the specific two-qubit interaction model under consideration.
Here we explore how such a restricted class of ``controlled unitaries'' can be characterized with high efficiency relative to  arbitrary two-qubit unitaries. Specifically, we study unitaries of the form 
\begin{equation}\label{eq:block-diag-u}
    U_\text{pulse} = \begin{pmatrix}
        e^{-i \phi} U_0 & 0 \\ 
        0 & e^{i \phi} U_1
    \end{pmatrix} \, , 
\end{equation}
where $U_0 = \exp(-i \, \mathbf{u} \cdot \mathbf{\sigma})$ and $U_1 = \exp(-i \, \mathbf{v} \cdot \mathbf{\sigma})$ are single-qubit unitaries, with $\sigma = (X, Y, Z)$. The superconducting cross-resonance pulse is one such interaction falling into this category, as we will show later. 

Characterizing a gate using this unitary structure involves estimating the value of 7 free parameters (compared to 15 in the general case), and enables a tomography scheme that is significantly more efficient and robust than the general case. Any unitary of this form has canonical coordinates $(c_1, 0, 0)$, with the closed-form formula for $c_1$ given in \cref{sec:app:ceff-derivation}.
Conversely, any unitary with canonical coordinates $(c_1, 0, 0)$ can be written in this form up to pre- and post-multiplication by two local unitaries $u_1u_1$ and $u_2u_2$, where $u_1$ and $u_2$ are constrained to rotation in the $XY$ plane (thus contributing 4 independent parameters).
Therefore, our method is applicable to tomography of any unitary where the required basis transformation to make it block-diagonal is known a priori (or can be determined through additional efficient tomography).

We show our characterization procedure for controlled pulses in \cref{fig:characterization}.
The first qubit controls which of these single-qubit unitaries is applied to the second qubit, and placing it in the $\ket{+}$ state allows us to perform `simultaneous' unitary tomography on both single-qubit Bloch vectors ($\mathbf{u}$ in $\ket{00}$ and $\ket{01}$, and $\mathbf{v}$ in $\ket{10}$ and $\ket{11}$) using a unified circuit configuration. Therefore, characterizing $U_\text{pulse}$ reduces to performing single-qubit gate tomography on the blocks, followed by estimating the phase difference between the blocks (giving $\phi$).

Supposing $U_0$ and $U_1$ have been characterized, one straightforward approach to subsequently estimate $\phi$ is as follows. Construct $U_2 = U_1^\dag U_0$, and select one of the two eigenstates $U_2 \ket{\psi} = e^{i \lambda} \ket{\psi}$. Then $U_1 \ket{\psi} = e^{-i \lambda} U_0 \ket{\psi}$. Therefore,
\begin{equation}
    U_\text{pulse} \ket{+}\ket{\psi} = \frac{1}{\sqrt{2}}(e^{-i \phi} \ket{0} + e^{i (\phi - \lambda)} \ket{1})U_0\ket{\psi} \, .
\end{equation}
Measuring the first qubit in the $X$ and $Y$ bases allows us to extract $\cos (\lambda - 2\phi)$ and $\sin (\lambda - 2\phi)$, giving a robust estimate of $\phi$. To enable curve fitting over different pulse iteration counts, the $U_0$ can be undone on the second qubit after each iteration, as shown in \cref{fig:characterization}.

\begin{figure*}[!t]
    \centering
    \includegraphics[width=\linewidth]{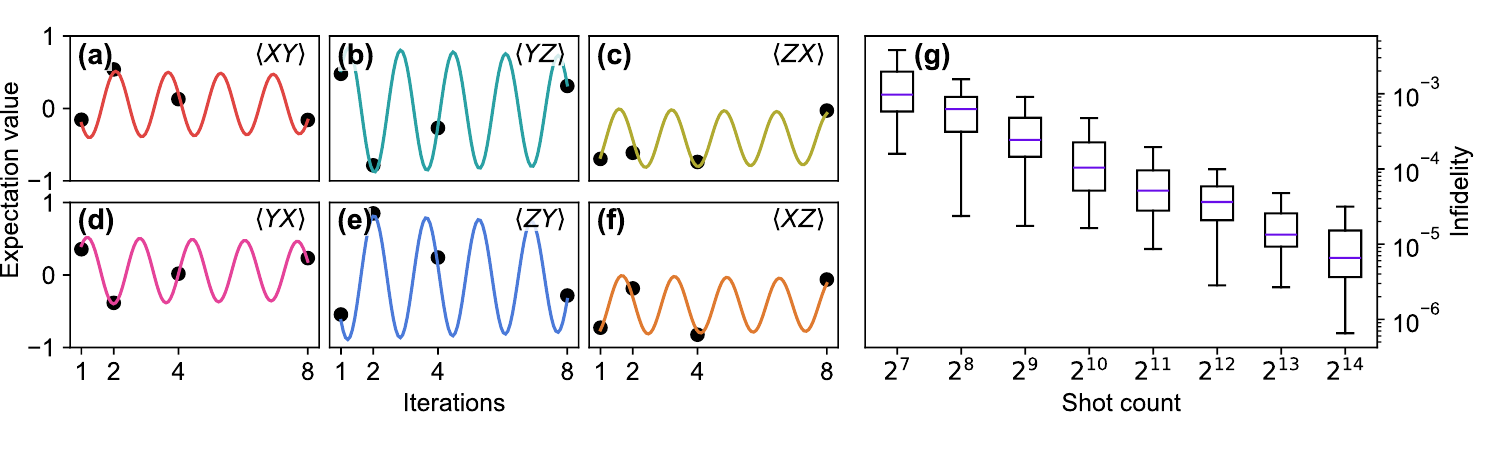}
    \caption{Numerical simulations of the controlled-pulse characterization process in the presence of shot noise and $10^{-2}$ depolarizing probability per application of $U_\text{pulse}$, and using a fixed iteration set of $\{1, 2, 4, 8\}$. \textbf{(a)-(f)} Example of the single-qubit tomography and fit data for one of two Haar-random $SU(2)$ matrices forming a block-diagonal unitary, using a shot count of 128 per experiment. The notation $\langle A B\rangle$ denotes state preparation in the $A$ basis and measurement in the $B$ basis. This example had a $U_\text{pulse}$ reconstruction infidelity of $1.4\times 10^{-4}$. \textbf{(g)} Box plot of the reconstruction infidelity (compared to the true unitary $U_\text{pulse}$) from the full fitting process to characterize $U_0$, $U_1$, and $\phi$. For each shot count, data was generated over 50 Haar-random controlled unitaries.}
    \label{fig:characterization-simulation}
\end{figure*}

This specialized characterization technique enables substantially fewer experiments than full two-qubit tomography for relevant interactions. Additionally, the characterization is subdivided into three individual optimizations on the data to estimate $\mathbf{u}$, $\mathbf{v}$ and $\phi$ separately. Each of these optimizations involve characterization of at most three parameters at once, rather than requiring fitting to all seven free parameters of a block-diagonal unitary simultaneously. We verify the stability of the approach in \cref{fig:characterization-simulation} via numerical simulations using the Q-CTRL Boulder Opal gradient-based optimizer, with a small iteration set of $\{1, 2, 4, 8\}$ and only 128 shots per circuit producing a reconstruction infidelity of approximately $10^{-3}$ on average.

\subsection{Accelerated compilation: single-axis pulses}
\label{sec:single-axis-pulse-compilation}

In this section we consider a restricted class of pulses where we can obtain closed-form synthesis conditions and gate parameters.
Compared to the generic optimization-based approach, having access to a symbolic circuit ansatz along with closed-form synthesis criteria in terms of the canonical coordinates of each pulse allows for faster pattern-based compilation. In addition, it is more amenable to symbolic compilation, where the gate parameters in a quantum circuit are left arbitrary at compilation time and assigned only at circuit runtime.

We define \textit{single-axis} pulses as those with canonical coordinates $\mathbf{c}=(c_\text{eff}, 0, 0)$.
The controlled pulses explored in the previous section have this property.
These pulses are locally equivalent to $e^{-i t P}$ for any two-qubit Pauli $P \in \{X, Y, Z\}^{\otimes 2}$.
Many modalities support controls to generate such gates, including the cross-resonance gate in superconducting systems.

We now show how to synthesize the canonical gate $\mathcal{C}(c_1, c_2, c_3)$ using the shortest-duration possible sequence of up to three characterized single-axis pulses, which enables synthesis of any unitary $U_\text{block}$ via the Cartan decomposition. It is sufficient to frame synthesis in terms of the canonical gates $\can(c_\text{eff}^{(i)}, 0, 0)$ associated with each characterized single-axis pulse $U_\text{pulse}^{(i)}$ since we can map between them using
\begin{equation}
    \can(c_\text{eff}^{(i)}, 0, 0) 
     = (w_1  w_2)^{-1} \cdot  U_\text{pulse}^{(i)} \cdot (w_3  w_4)^{-1} \, ,
    \label{eq:single-axis-pulse-kak-decomposition}
\end{equation}
via the Cartan decomposition. With this in mind, we consider the three cases for the number of nonzero canonical coordinates of an input circuit block, respectively requiring 1-2, 2, and 3 pulses to synthesize.

\subsubsection{Case 1: $U_\text{block} \sim \can(c_1, 0, 0)$}

If there exists a characterized pulse index $i$ in the extended gate set such that $c_\text{eff}^{(i)}=c_1$, then (trivially) just one pulse is required. Otherwise, two pulses are required. In Case 2, we show how to implement any $(c_1, c_2, 0)$ optimally using two pulses, and thus handle $(c_1, 0, 0)$ via Case 2.

\subsubsection{Case 2: $U_\text{block} \sim \can(c_1, c_2, 0)$}

For each pair of characterized pulse indices $\mathbf{k}=(k_1, k_2)$, define the circuit motif
\begin{equation}
 V_\mathbf{k}(\phi)=\can(c_\text{eff}^{(k_1)}, 0, 0) \cdot (Z(\phi_1)  Z(\phi_2)) \cdot \can(c_\text{eff}^{(k_2)}, 0, 0) \, ,   
 \label{eq:ansatz-c1c20}
\end{equation}
assuming without loss of generality that $c_\text{eff}^{(k_1)} \geq c_\text{eff}^{(k_2)}$.
It is sufficient to restrict to single-qubit $Z$ rotations in this case, since a full $X(\alpha_i)Z(\phi_i)X(\beta_i)$ parameterization of $SU(2)$ would have the two outer rotations commute through the two-qubit canonical gates, therefore leaving the invariants unchanged.
Algebraically calculating the Makhlin invariants and solving for
\begin{equation}
    \label{eq:makhlin-equivalence-single-axis-pulse}
    \mathbf{g}(\can(c_1, c_2, 0)) = \mathbf{g}(V_\mathbf{k}(\phi)) \, ,
\end{equation}
we find the necessary closed-form $\phi$ expressions through directly solving this system of equations, and we provide them in \cref{app: single parameter identities}.
The conditions for these angles to be real-valued, and thus for the synthesis to be achievable for a given pair of pulses, is
\begin{align}
    \label{eq:conditions-c1c20}
    c_\text{eff}^{(1)} + c_\text{eff}^{(2)} &\geq c_1 + c_2 \, ,\\
    c_\text{eff}^{(1)} - c_\text{eff}^{(2)} &\leq c_1 - c_2 \, .
\end{align}
This is consistent with the conditions found in~\cite{Zhang2005,Peterson2022}.
To find the outer unitaries to satisfy
\begin{equation}
    \can(c_1, c_2, 0) = (Z(\theta_1)  Z(\theta_2)) \cdot V_\mathbf{k}(\phi) \cdot (Z(\theta_3)  Z(\theta_4)) \, ,
    \label{eq:motif-c1-c2-0}
\end{equation}
we algebraically solve the system of equations from matching matrix elements, and find the necessary $\theta$ in closed form (\cref{app: single parameter identities}). Observe that also restricting the outer unitaries to $Z$ rotations is sufficient to achieve the desired synthesis.

This ansatz is sufficient to optimally synthesize \textit{any} unitary locally equivalent to $\can(c_1, c_2, 0)$ using two single-axis pulses.
We simply iterate through pairs of pulses in ascending order of total duration, choose the first pair that satisfies the two synthesis inequalities, and substitute the appropriate angle values.
Finally, we back-substitute \cref{eq:ansatz-c1c20} to obtain a circuit for $\can(c_1, c_2, 0)$ in terms of characterized pulses and single-qubit gates, and then use the Cartan decomposition to turn the synthesis for $\can(c_1, c_2, 0)$ into a synthesis for $U_\text{block}$.

\subsubsection{Case 3: $U_\text{block} \sim \can(c_1, c_2, c_3)$}

For the remaining case of a general $\mathcal{C}(c_1, c_2, c_3)$ block using a total of three pulses, we can apply the two-pulse motif of Case 2 twice. First, rewrite $\mathcal{C}(c_1, c_2, c_3) = \mathcal{C}(c_1, c_2, 0) \cdot \mathcal{C}(0, 0, c_3)$, using the identity $\can(\bm{c} + \bm{c}') = \can(\bm{c}) \cdot \can(\bm{c}')$. Then, use the Case 2 motif to synthesize $\can(c_1, c_2, 0)$ into $\can(c_\text{eff}^{(k_1)}, 0, 0)$ and $\can(\delta, 0, 0)$ for an intermediate/``temporary'' interaction strength $\delta$, which will be replaced by one of our ``real'' available interaction strengths in the subsequent motif step. Due to the local $Z$ rotations, the $\mathcal{C}(0, 0, c_3)$ gate can commute next to the $\mathcal{C}(\delta, 0, 0)$ gate. Through local operations, we can simplify to a $\mathcal{C}(\delta, c_3, 0)$ gate.
Then the same Case 2 motif can be applied again, transforming $\can(\delta, c_3, 0)$ into $\can(c_\text{eff}^{(k_2)}, 0, 0)$ and $\can(c_\text{eff}^{(k_3)}, 0, 0)$. This construction generates a synthesis of $\mathcal{C}(c_1, c_2, c_3)$ into three $\can(c_\text{eff}^{(k_i)}, 0, 0)$ gates along with local unitaries, which can be used to synthesize an arbitrary $U_\text{block}$ in the same way as above.

The repeated application of the \cref{eq:ansatz-c1c20} motif gives a set of inequalities that the intermediate value $\delta$ must satisfy. The range of values $\delta$ can take is non-empty exactly when
\begin{align}
    c_\text{eff}^{(k_1)} + c_\text{eff}^{(k_2)} + c_\text{eff}^{(k_3)} &\geq c_1 + c_2 + c_3 \, , \\
    -c_\text{eff}^{(k_1)} + c_\text{eff}^{(k_2)} + c_\text{eff}^{(k_3)} &\geq -c_1 + c_2 + c_3 \, , \\
    c_\text{eff}^{(k_3)} &\geq c_3 \, .
    \label{eq:c1-c2-c3-synthesis-conditions-single-axis}
\end{align}
We prove this result in \cref{app: single parameter identities}.
This matches the polytope in \cite{Peterson2022}, demonstrating optimality.
Furthermore, our motif is consistent with the analogous constructive method for $XX$ pulses which is presented in that work.

Overall, we have shown how to synthesize any two-qubit block using up to three single-axis pulses.
The circuit structures are fixed for each case, with the relevant synthesis conditions and gate parameters expressed in closed-form.
The worst-case situation is a $SWAP$ gate, which has coordinates $\mathbf{c} = (\pi/2, \pi/2, \pi/2)$ and requires three $(\pi/2, 0, 0)$ pulses to implement.
This method enables a highly efficient pulse compilation protocol for on relevant platforms.

\subsection{Example application: the cross-resonance interaction}
\label{sec:cross-resonance-characterization}

\begin{figure}[!t]
    \centering
    \includegraphics[width=\linewidth]{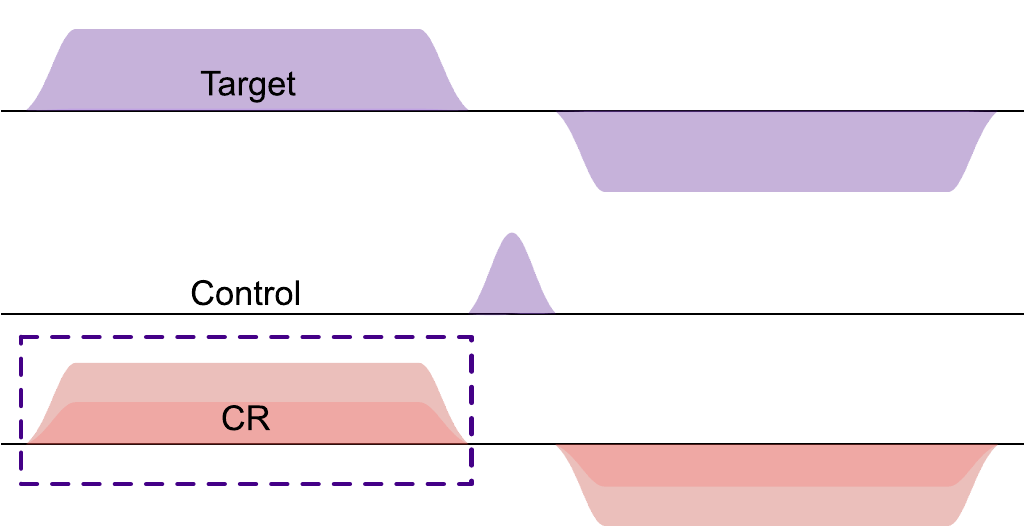}
    \caption{Illustration of the waveform for a calibrated echoed cross-resonance gate. The full waveform implements a gate locally equivalent to $CX$.
    The raw cross-resonance waveform in the dashed box can be extracted and characterized to obtain a gate locally equivalent to $\can(\pi/4 + \delta, 0, 0)$, for some (likely small) value $\delta$ which is determined through the characterization process. The area of the raw CR waveform can also be scaled by a factor $\alpha$ to obtain, through characterization, a gate locally equivalent to $\can(\theta_\alpha, 0, 0)$ with $\theta_\alpha = \alpha \pi/4 + \delta_\alpha$. This allows approximate targeting of desirable additional angles for an extended gate set.}
    \label{fig:ecr-pulse-schedule}
\end{figure}

The superconducting cross-resonance pulse is an example interaction where both the controlled-pulse characterization method and the single-axis compilation method can be applied.
The Hamiltonian~\cite{Magesan2020} for a cross-resonance pulse is modeled as
\begin{align}\label{effective Hamiltonian}
    H_{\text{cr}} &= \frac{1}{2} \left( \nu_{zx}ZX +\nu_{zy}ZY +\nu_{zz}ZZ \right. \\
    \nonumber
    & \qquad \left. +\nu_{ix}IX +\nu_{iy}IY +\nu_{iz}IZ + \nu_{zi}ZI \right) \, .
\end{align}
The coefficients depend nonlinearly on the control parameters, such as the drive amplitude.
The $\nu_{ZX}$ term is dominant~\cite{Chow2011}, and thus the typical calibration target is to suppress all other coefficients and obtain a pure $ZX(\pi/2)$ rotation, which is locally equivalent to $CX$. An echoed cross-resonance (ECR) pattern~\cite{Sheldon2016}, illustrated in \cref{fig:ecr-pulse-schedule}, is typically utilized to cancel out terms at the circuit level, along with a calibrated tone on the target to eliminate other error terms.
This sequence leads to
\begin{align}\label{echoed cr arbitrary angle defined}
    ZX(\frac{\pi}{2}) \approx \underbrace{e^{-i H_\text{cr}} \cdot \left(X  I\right) \cdot  e^{i H_\text{cr}}}_{\text{ECR}}  \cdot \left(X  I\right) \, .
\end{align}

Suppose we extract, optionally rescale, and characterize the raw cross-resonance pulse $U_\text{pulse} = \exp({- i H_\text{cr}})$ from the $ECR$ calibration, shown as the dashed region in \cref{fig:ecr-pulse-schedule}. Rather than treating the non-$ZX$ terms in \cref{effective Hamiltonian} as error terms which must be suppressed, our motivation is instead to absorb these terms into the gate definition.
Hence, there is no requirement for an echo pattern.

Equating the direct calculation of the matrix corresponding to $e^{-i H_\text{cr}}$ with \cref{eq:block-diag-u} gives
\begin{equation}
    \mathbf{u} = \begin{pmatrix}
        \nu_{zx} + \nu_{ix} \\
        \nu_{zy} + \nu_{iy} \\
        \nu_{zz} + \nu_{iz}
    \end{pmatrix} \, ,
    \qquad
    \mathbf{v} = \begin{pmatrix}
        \nu_{zx} - \nu_{ix} \\
        \nu_{zy} - \nu_{iy} \\
        \nu_{zz} - \nu_{iz}
    \end{pmatrix} \, ,
\end{equation}
and $\nu_{zi} = \phi$.
Therefore, the cross-resonance interaction is a controlled pulse, and we can utilize the efficient controlled-pulse characterization process along with the single-axis pulse compilation process.
All terms except $\nu_{zi}$ contribute to the canonical coordinate $(c_1, 0, 0)$ as per \cref{sec:app:ceff-derivation}, and thus even the single-qubit Hamiltonian coefficients that are traditionally considered ``errors'' in fact give us additional entangling power in the characterized pulses.
In a sense, calibration protocols that eliminate these terms are using additional resources to subtract from the power of a gate, while our characterize-and-compile approach embraces the addition to the overall gate entangling power (via increasing the canonical coordinate $c_1$) that these terms contribute. 

\section{Algorithmic Benchmarking}\label{sec: Experimental Algorithmic Benchmarking}

\begin{figure*}[ht!]
    \centering
    \hfill
    \begin{minipage}[c]{0.33\linewidth}
    \includegraphics[width=\linewidth]{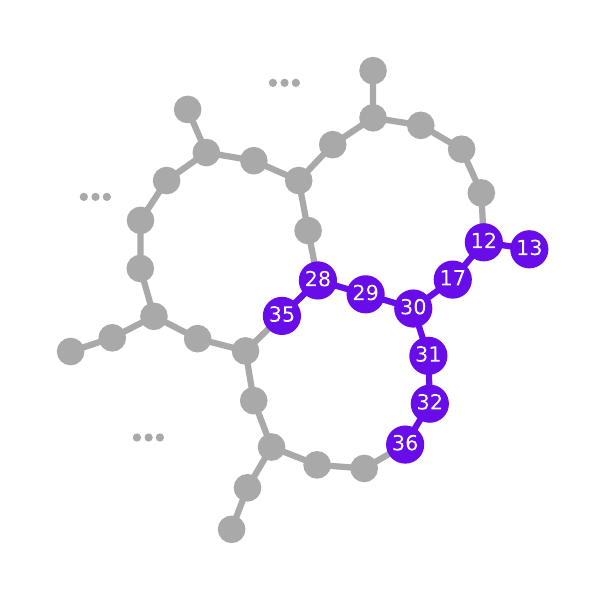}
    \end{minipage}
    \begin{minipage}[c]{0.65\linewidth} \centering
    \begin{tabular}[b]{|r|r|r|r|r|r|r|r|r|}
    \hline
    \multicolumn{1}{|c|}{\multirow{2}{*}{\textbf{Pair}}} & \multicolumn{7}{c|}{\textbf{Characterized Hamiltonian coefficients}} & \multirow{2}{*}{\textbf{Weyl coord.}} \\ \cline{2-8}
     & ZX & ZY & ZZ & IX & IY & IZ & ZI &  \\ \hline
(12, 13) & $0.827$ & $-0.020$ & $-0.047$ & $0.687$ & $-0.028$ & $0.055$ & $2.352$ & $\pi/4 +  0.043 $\\ \hline
(12, 17) & $0.845$ & $0.018$ & $-0.009$ & $-0.423$ & $-0.023$ & $0.042$ & $-2.801$ & $\pi/4 +  0.060 $\\ \hline
(17, 30) & $0.832$ & $0.168$ & $-0.006$ & $-0.260$ & $-0.073$ & $0.004$ & $-2.899$ & $\pi/4 +  0.064 $\\ \hline
(28, 29) & $0.854$ & $0.046$ & $0.036$ & $0.072$ & $0.029$ & $-0.025$ & $2.106$ & $\pi/4 +  0.070 $\\ \hline
(28, 35) & $0.836$ & $-0.005$ & $-0.008$ & $-0.291$ & $-0.024$ & $0.003$ & $3.088$ & $\pi/4 +  0.051 $\\ \hline
(29, 30) & $0.843$ & $0.000$ & $-0.009$ & $0.048$ & $0.006$ & $-0.019$ & $-2.905$ & $\pi/4 +  0.058 $\\ \hline
(30, 31) & $0.836$ & $-0.012$ & $-0.032$ & $-0.073$ & $-0.019$ & $0.011$ & $2.695$ & $\pi/4 +  0.051 $\\ \hline
(31, 32) & $0.839$ & $-0.044$ & $-0.047$ & $-0.014$ & $-0.029$ & $0.068$ & $-0.879$ & $\pi/4 +  0.056 $\\ \hline
(32, 36) & $0.830$ & $0.155$ & $0.006$ & $-0.462$ & $-0.105$ & $-0.025$ & $-3.109$ & $\pi/4 +  0.059 $\\ \hline
    \end{tabular}
    \end{minipage}
    \hfill
    \caption{An example of a layout for benchmarking a 10-qubit Quantum Fourier Transform circuit on \texttt{ibm\_brisbane}, and a characterized efficient pulse gate for each physical qubit pair involved in the circuit. The table shows the characterized Hamiltonian coefficients $\nu$ of \cref{effective Hamiltonian} along with the Weyl coordinate $(c_1, 0, 0)$ which is associated with the pulse $U_\text{pulse}=e^{-i H_\text{cr}}$ for each qubit pair. This characterized pulse is turned into a pure $\can(c_1, 0, 0)$ gate via single-qubit rotations determined by the Cartan decomposition, for later use by the compiler.}
    \label{fig:characterization-data-10q-qft}
\end{figure*}

As an example of the impact of our characterize-and-compile methodology, in this section we show how adding just one efficient pulse gate to the basis set can improve the hardware performance of two quantum algorithms, the Quantum Fourier Transform and Trotter simulation. Algorithmic benchmarking is the most straightforward and meaningful way to show practical benefit of compilation and quantum control techniques~\cite{Lubinski2023}.

Our experiments are conducted on IBM's 127-qubit superconducting device \texttt{ibm\_brisbane}, having ``default'' gate set $\{Z(\theta), \sqrt{X}, X, ECR\}$. We obtain a single efficient gate for physical qubit pairs by extracting the raw CR waveform as illustrated in \cref{fig:ecr-pulse-schedule}, using the Qiskit Pulse API\footnote{IBM has since removed the Qiskit Pulse API, replacing it with calibrated parameterized (``fractional'') two-qubit gates~\cite{Almeida2024}.}~\cite{Alexander2020}. We scale the rectangular area of the extracted waveforms by a small amount to target an angle of $\pi/4 + 0.05$, meaning that with high probability the final characterized angle would be at least $\pi/4$. This minimum angle ensures that two efficient pulses are sufficient to reach every point on the $(c_1, 0, 0)$ line of the Weyl chamber (per \cref{eq:conditions-c1c20}) without needing to resort to an $ECR$ gate.
We also incorporate a calibrated cancellation/rotary tone on the target qubit based on the calibrated waveform, converting it to an equivalent Gaussian square echo \cite{Sundaresan2020,Jurcevic2021} to induce partial cancellation of the non-dominant cross-resonance terms in \cref{effective Hamiltonian}. The addition of this cancellation tone is not strictly necessary with respect to our characterization approach, since all characterized Hamiltonian terms are accounted for in software irrespective of their magnitude, but is convenient to increase the speed of optimization convergence via improved initial guesses (of 0) for the associated Hamiltonian terms.

We perform characterization immediately before each set of algorithmic benchmarking experiments, meaning drift is not a consideration in our results.
We use an iteration set of $\{1, 2, \ldots, 13\}$ for each characterization, and 4096 shots per characterization circuit.
Our characterization for multiple two-qubit pairs is performed in parallel, while taking care to ensure no gates involving shared qubits are characterized in the same batch.
This minimizes the number of separate circuit executions required to perform large-scale characterization across a device.
The optimal number of independent experimental batches that are required to characterize an efficient gate for each qubit pair corresponds to the chromatic index of the device topology graph.
On the heavy-hex topology of \texttt{ibm\_brisbane} and other IBM devices, this requires just three independent batches.

For each algorithmic benchmarking experiment, we ensure that the circuits differ only in the available gate sets, comparing performance with and without the characterized additional gate $\can(\pi/4 + \delta, 0, 0)$ on involved qubit pairs. Examples of the $\delta$ values obtained are displayed in \cref{fig:characterization-data-10q-qft}, for the set of qubit pairs used to execute a particular 10-qubit QFT circuit. In our algorithmic benchmarking, all circuits are executed on the same qubit layouts, with the same number of shots, and with identical execution settings. Measurement error mitigation is applied to the raw counts for all circuits.

\subsection{Quantum Fourier Transform}

\begin{figure}[ht!]
    \centering
    \includegraphics[width=1.0\linewidth]{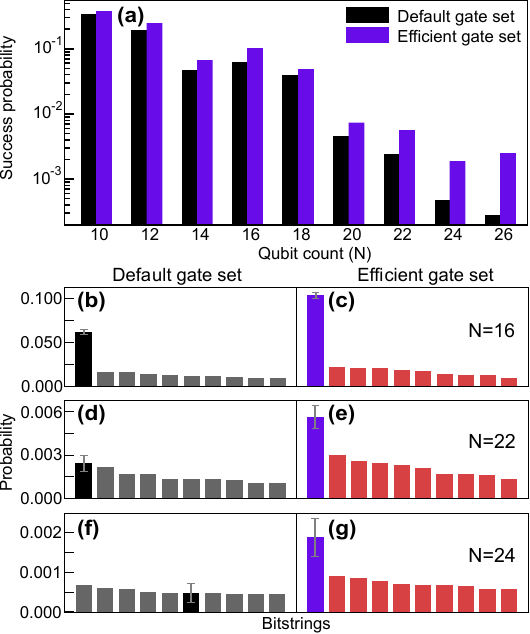} 
    \caption{Experimental comparison of inverse quantum Fourier transform circuits run on \texttt{ibm\_brisbane}, with identical execution workflows except for the usage of efficient gates in addition to backend default gates versus usage of only backend default gates. The target bitstring is $101010\dots10$.  \textbf{(a)} Observed likelihood of obtaining the target bitstring as a function of circuit width. \textbf{(b)-(g)} Histograms of observed results for three selected circuit widths $N=16, 22, 24$ respectively. Only the 10 most frequently observed bitstrings are depicted. In all cases, the success probability is substantially boosted by using efficient gates. For the 24-qubit QFT, the use of efficient gates allows the correct result to emerge above the noise as the most likely result. Standard error bars on the target bitstring represent uncertainty due to shot noise.}
    \label{fig:inv qft} 
\end{figure}

The first algorithmic benchmark we employ is the Quantum Fourier Transform (QFT).  This is an algorithm where benefit from efficient gates is expected, because it is primarily made up of controlled-phase gates of different rotation angles (along with swap gates when compiling onto the limited-connectivity heavy-hex topology of \texttt{ibm\_brisbane}). Controlled-phase gates have canonical coordinates $(c_1, 0, 0)$, enabling efficient synthesis using at most two applications of the pulse-efficient gate.

For the purposes of benchmarking performance, we utilize the inverse QFT circuit primitive.
We apply this primitive to an initial state of tailored single-qubit rotations.
These single-qubit rotations are chosen such that the ideal output is a computational basis state, a choice which enables a straightforward evaluation metric: the probability of obtaining this target bitstring.

We conduct experiments for even circuit widths (qubit numbers, $N$) ranging from $N=10$ to $N=26$. For the target bitstrings, we choose the all-zeros bitstring, the all-ones bitstring, alternating 0101...01 and 1010...10 bitstrings, and the double-alternating bitstring 0011...0011, to average out bias with respect to any particular noise channel. Each experiment is repeated five times using 8000 shots.

The QFT experimental benchmarking results are shown in \cref{fig:inv qft}, comparing the probability of obtaining the target bitstring with and without compiler utilization of the characterized efficient entangling gate. We show the probability of obtaining the target bitstring, averaged over all experiments, for each qubit count. Notably, there is increasing advantage from utilization of this single additional entangling gate as the size of the QFT circuit increases. The comparison to the default gate set across different qubit counts, shown in \cref{fig:inv qft} (a), shows up to a $7$X improvement in the algorithmic success probability for up to $N=26$ qubits. For three selected qubit counts, we also show the histogram of the (odd-even target) output distribution, limited to the 10 most-sampled bitstrings. Notably, in the $N=24$ case, the default gate set does not achieve the target bitstring as the most probable. In contrast, the inclusion of the single additional efficient gate boosts the fidelity of the circuit to the extent that the target bitstring becomes the mode of the output distribution.

\subsection{Hamiltonian Simulation}

\begin{figure*}[!ht]
    \centering
    \includegraphics[width=1.0\textwidth]{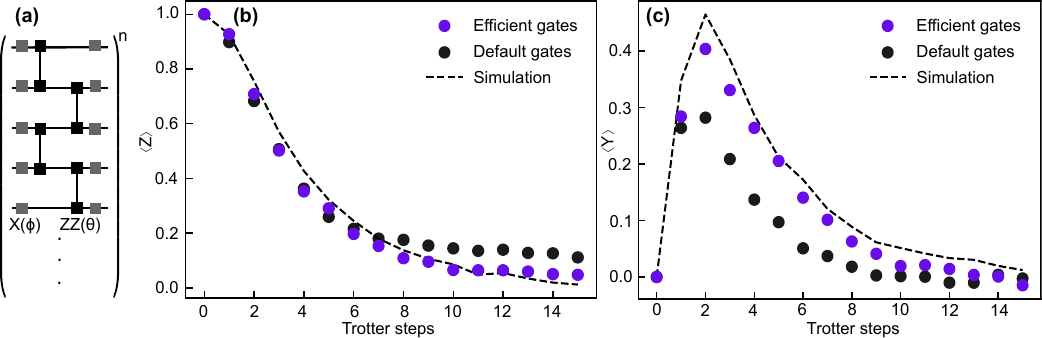}
    \caption{
    \label{fig:trotter tfim}
    Experimental comparison of Hamiltonian simulation of the TFIM with default versus efficient gates. The implementation uses $N=25$ qubits connected on a one-dimensional line. {\bf{(a)}} A single Trotter layer. The ZZ($\theta$) gates are synthesized using two efficient gates locally equivalent to $\can(\pi/4, 0, 0)$, in contrast with two ECR gates in the default case. {\bf{(b)}} Observed expectation value of the average $Z$ magnetization. {\bf{(c)}} Observed expectation value of the average $Y$ magnetization. The resulting trajectories using efficient gates are significantly closer to the ideal simulation of the Trotterized evolution compared to the circuits that use only default gates with a 9X smaller MSE for $\langle Y \rangle$ and a 5X smaller MSE for $\langle Z \rangle$). 
    }
\end{figure*}

We simulate the one-dimensional transverse field Ising model with Hamiltonian
\begin{align}
    H = -J \sum_i Z_i Z_{i+1} + h \sum_i X_i \, .
\end{align}
Here $J$ is the coupling between nearest neighboring spins and $h$ is the global transverse magnetic field strength.
We use the second-order Trotter-Suzuki product formula to simulate the evolution for the full time duration $t = n \, dt$ by breaking it into $n$ Trotter steps.
\cref{fig:trotter tfim} (a) shows a circuit schematic for each Trotter step.

We implement the transverse field Ising model Hamiltonian with $N=25$ qubits connected on a one-dimensional line topology, with parameters $J=h=1$. Each circuit represents an evolution for time $t=n \, dt$, where $dt = \pi/15$ and $n$ is the number of Trotter steps.

For this one-dimensional line topology, linear nearest-neighbor qubit connectivity is sufficient to compile these circuits without requiring swap gates, meaning that the only two-qubit interactions are $ZZ$ rotations. This makes these Trotter circuits amenable to hardware performance improvement from efficient gates. Without efficient gates, the compiled circuits require a total of $48n$ $ECR$ gates (two per ZZ$(\theta)$ operation). With characterized efficient gates $(\pi/4 + \delta, 0, 0)$, every $ECR$ in the circuit is replaced by an efficient gate.

We execute each Trotter step circuit using 4096 shots. In \cref{fig:trotter tfim}, we plot the obtained expectation value of the average $Y$ and $Z$ magnetization with and without the use of a single efficient gate on each qubit pair, comparing to an ideal (error-free) classical simulation.
For both the $Y$ and $Z$ magnetizations, efficient gates produce curves that match the expected outcome from ideal simulation significantly better than the default implementation.
The MSE is $9X$ smaller with efficient gates compared to $ECR$ gates for $\langle{Z}\rangle$. We emphasize that there is no error mitigation employed in these experiments, meaning no additional overheads are required through the use of efficient gates in executing the Trotter experiments.

The significant improvements obtained from the addition of a single efficient gate to each qubit pair in an algorithm demonstrates the practical viability of our method in enhancing algorithmic performance on quantum hardware with minimal calibration overhead.

\section{Summary and Discussion}
\label{sec:summary}

In this work, we have presented an automated and resource-efficient technique to extend the entangling-gate set of a quantum device with arbitrary pulses.
We set out a compilation scheme that leverages these gates, leading to circuits with reduced duration which accumulate less coherent error and are as a result executed with higher fidelity.

Our approach is inspired by a shift in perspective from detailed and high-fidelity calibration to characterization.
Instead of calibrating analog-layer control pulses to achieve additional gates, or performing challenging calibration of a continuously parameterized entangling interaction, we focus on accurately and efficiently characterizing the unitary realized by an essentially arbitrary waveform, then utilizing this information and various numerical and analytical techniques in circuit-level compilation.
By avoiding extensive online closed-loop gate calibration, our approach is scalable and practical for supplementing large devices with efficient gates, while ensuring high-fidelity gate and algorithm performance.
The overall performance of this approach is limited by the extent to which the gate errors are describable as unitary, meaning incoherent energy relaxation processes are not addressed.

The structural assumptions underlying our method dramatically reduce the characterization burden relative to more comprehensive, and hence unscalable, approaches.
We leverage the assumption of unitarity to reduce the characterization burden; this accelerated tomography enables improved performance compared to calibration of specific exact target points in the Weyl chamber due to the elimination of the closed-loop optimization feedback loop.
We further assume access to a special perfect entangler in the original gate set for each qubit pair, which guarantees reachability of the full $SU(4)$ space with at most three non-local gates.

Using our compiler synthesis technique to utilize efficient gates, we benchmark a standard algorithm, the inverse Quantum Fourier Transform (QFT)~\cite{Lubinski2023}, and compare to performance without the additional efficient gates.
The procedure demonstrates up to a $7$X improvement in the algorithmic success probability for up to $N=26$ qubits. We also test Trotterized Hamiltonian simulation applied to the one-dimensional transverse-field Ising model on a line~\cite{Sawaya2024}, using $25$ qubits and up to $15$ steps.
We observe an improvement of up to $9$X in the MSE in the calculation of expectation values of observables.

Our techniques are hardware-agnostic: the characterization and compilation techniques apply to completely arbitrary two-qubit pulses on any platform.
Given a particular architecture, our results can be used both as a direct characterization and compilation layer and, if desired, to inform the design of further accelerated characterization and compilation approaches tailored for the specific modality, as exemplified by the two specific cases of block-diagonal pulses and single-axis pulses explored in this work.

Several future research directions follow naturally from this study.
A key question is the temporal stability of characterized pulses. Analyzing how long a given gate definition remains valid under drift, and how frequently re-characterization is required, is essential for large-scale deployment.
Another avenue is refining strategies for calibrating the special perfect entangling gate itself, for example by optimizing a loss defined on the Makhlin invariants rather than a specific target unitary.
This can be combined with our other optimization techniques for ensuring high-fidelity and drift-robust gates~\cite{Ball2021}.
Our compilation framework also suggests approximate two-qubit gate synthesis schemes that trade small unitary approximation error for reduced block duration, by minimizing differences between Makhlin invariants even when a block's Weyl coordinates lie outside the polytope generated by a given efficient gate sequence.
Relatedly, recent work by \citet{McKinney2025} demonstrates an alternative two-qubit synthesis strategy via monodromy techniques, determining the necessary inner single-qubit gate parameters by linear programming. Integrating such a method with our characterization-driven pipeline is an interesting direction.
Finally, extending the closed-form motifs for single-axis pulses to support ``multi-axis'' pulses, such as M{\o}lmer-S{\o}rensen pulses on ion-trap architectures, would further reduce compilation overhead for these modalities.

Overall, this study presents new techniques enabling execution of quantum circuits faster and with higher fidelity, requiring modest characterization overheads and avoiding substantial calibration overheads.
Our techniques apply irrespective of the quantum hardware platform, and showcase that significant performance gains can be achieved via solutions purely at the software level through co-design of characterization techniques and compiler architecture.

\section{Acknowledgements}

We are grateful to all other colleagues at Q-CTRL whose technical, product engineering, and design work has supported the results presented in this paper.

\bibliography{cite.bib}

@article{DiVincenzo1995,
  title = {Two-bit gates are universal for quantum computation},
  author = {DiVincenzo, David P.},
  journal = {Phys. Rev. A},
  volume = {51},
  issue = {2},
  pages = {1015--1022},
  numpages = {0},
  year = {1995},
  month = {Feb},
  publisher = {American Physical Society},
  doi = {10.1103/PhysRevA.51.1015},
  url = {https://link.aps.org/doi/10.1103/PhysRevA.51.1015}
}

@article{Lloyd1995,
  title = {Almost Any Quantum Logic Gate is Universal},
  author = {Lloyd, Seth},
  journal = {Phys. Rev. Lett.},
  volume = {75},
  issue = {2},
  pages = {346--349},
  numpages = {0},
  year = {1995},
  month = {Jul},
  publisher = {American Physical Society},
  doi = {10.1103/PhysRevLett.75.346},
  url = {https://link.aps.org/doi/10.1103/PhysRevLett.75.346}
}

@article{Kitaev1997,
  title = {Quantum computations: algorithms and error correction},
  volume = {52},
  ISSN = {1468-4829},
  url = {http://dx.doi.org/10.1070/RM1997v052n06ABEH002155},
  DOI = {10.1070/rm1997v052n06abeh002155},
  number = {6},
  journal = {Russ. Math. Surv.},
  publisher = {Steklov Mathematical Institute},
  author = {Kitaev,  A Yu},
  year = {1997},
  month = dec,
  pages = {1191–1249}
}

@article{Majer2007,
  title = {Coupling superconducting qubits via a cavity bus},
  volume = {449},
  ISSN = {1476-4687},
  url = {http://dx.doi.org/10.1038/nature06184},
  DOI = {10.1038/nature06184},
  number = {7161},
  journal = {Nature},
  publisher = {Springer Science and Business Media LLC},
  author = {Majer,  J. and Chow,  J. M. and Gambetta,  J. M. and Koch,  Jens and Johnson,  B. R. and Schreier,  J. A. and Frunzio,  L. and Schuster,  D. I. and Houck,  A. A. and Wallraff,  A. and Blais,  A. and Devoret,  M. H. and Girvin,  S. M. and Schoelkopf,  R. J.},
  year = {2007},
  month = sep,
  pages = {443–447}
}

@article{DiCarlo2009,
  title = {Demonstration of two-qubit algorithms with a superconducting quantum processor},
  volume = {460},
  ISSN = {1476-4687},
  url = {http://dx.doi.org/10.1038/nature08121},
  DOI = {10.1038/nature08121},
  number = {7252},
  journal = {Nature},
  publisher = {Springer Science and Business Media LLC},
  author = {DiCarlo,  L. and Chow,  J. M. and Gambetta,  J. M. and Bishop,  Lev S. and Johnson,  B. R. and Schuster,  D. I. and Majer,  J. and Blais,  A. and Frunzio,  L. and Girvin,  S. M. and Schoelkopf,  R. J.},
  year = {2009},
  month = jun,
  pages = {240–244}
}

@article{Rigetti2010,
  title = {Fully microwave-tunable universal gates in superconducting qubits with linear couplings and fixed transition frequencies},
  author = {Rigetti, Chad and Devoret, Michel},
  journal = {Phys. Rev. B},
  volume = {81},
  issue = {13},
  pages = {134507},
  numpages = {7},
  year = {2010},
  month = {Apr},
  publisher = {American Physical Society},
  doi = {10.1103/PhysRevB.81.134507},
  url = {https://link.aps.org/doi/10.1103/PhysRevB.81.134507}
}

@article{Chow2013,
  title = {Microwave-activated conditional-phase gate for superconducting qubits},
  volume = {15},
  ISSN = {1367-2630},
  url = {http://dx.doi.org/10.1088/1367-2630/15/11/115012},
  DOI = {10.1088/1367-2630/15/11/115012},
  number = {11},
  journal = {New J. Phys.},
  publisher = {IOP Publishing},
  author = {Chow,  Jerry M and Gambetta,  Jay M and Cross,  Andrew W and Merkel,  Seth T and Rigetti,  Chad and Steffen,  M},
  year = {2013},
  month = nov,
  pages = {115012}
}

@article{Casparis2019,
  title = {Voltage-controlled superconducting quantum bus},
  author = {Casparis, L. and Pearson, N. J. and Kringh\o{}j, A. and Larsen, T. W. and Kuemmeth, F. and Nyg\aa{}rd, J. and Krogstrup, P. and Petersson, K. D. and Marcus, C. M.},
  journal = {Phys. Rev. B},
  volume = {99},
  issue = {8},
  pages = {085434},
  numpages = {7},
  year = {2019},
  month = {Feb},
  publisher = {American Physical Society},
  doi = {10.1103/PhysRevB.99.085434},
  url = {https://link.aps.org/doi/10.1103/PhysRevB.99.085434}
}

@article{Wei2022,
  title = {Hamiltonian Engineering with Multicolor Drives for Fast Entangling Gates and Quantum Crosstalk Cancellation},
  author = {Wei, K. X. and others},
  journal = {Phys. Rev. Lett.},
  volume = {129},
  issue = {6},
  pages = {060501},
  numpages = {6},
  year = {2022},
  month = {Aug},
  publisher = {American Physical Society},
  doi = {10.1103/PhysRevLett.129.060501},
  url = {https://link.aps.org/doi/10.1103/PhysRevLett.129.060501}
}

@article{Abrams2020,
  title = {Implementation of XY entangling gates with a single calibrated pulse},
  volume = {3},
  ISSN = {2520-1131},
  url = {http://dx.doi.org/10.1038/s41928-020-00498-1},
  DOI = {10.1038/s41928-020-00498-1},
  number = {12},
  journal = {Nature Electronics},
  publisher = {Springer Science and Business Media LLC},
  author = {Abrams,  Deanna M. and Didier,  Nicolas and Johnson,  Blake R. and Silva,  Marcus P. da and Ryan,  Colm A.},
  year = {2020},
  month = nov,
  pages = {744–750}
}

@article{Foxen2020,
  title = {Demonstrating a Continuous Set of Two-Qubit Gates for Near-Term Quantum Algorithms},
  author = {Foxen, B. and others},
  collaboration = {Google AI Quantum},
  journal = {Phys. Rev. Lett.},
  volume = {125},
  issue = {12},
  pages = {120504},
  numpages = {6},
  year = {2020},
  month = {Sep},
  publisher = {American Physical Society},
  doi = {10.1103/PhysRevLett.125.120504},
  url = {https://link.aps.org/doi/10.1103/PhysRevLett.125.120504}
}

@article{Peterson2020,
  title = {Two-Qubit Circuit Depth and the Monodromy Polytope},
  volume = {4},
  ISSN = {2521-327X},
  url = {http://dx.doi.org/10.22331/q-2020-03-26-247},
  DOI = {10.22331/q-2020-03-26-247},
  journal = {Quantum},
  publisher = {Verein zur Forderung des Open Access Publizierens in den Quantenwissenschaften},
  author = {Peterson,  Eric C. and Crooks,  Gavin E. and Smith,  Robert S.},
  year = {2020},
  month = mar,
  pages = {247}
}

@article{Collodo2020,
  title = {Implementation of Conditional Phase Gates Based on Tunable $ZZ$ Interactions},
  author = {Collodo, Michele C. and Herrmann, Johannes and Lacroix, Nathan and Andersen, Christian Kraglund and Remm, Ants and Lazar, Stefania and Besse, Jean-Claude and Walter, Theo and Wallraff, Andreas and Eichler, Christopher},
  journal = {Phys. Rev. Lett.},
  volume = {125},
  issue = {24},
  pages = {240502},
  numpages = {5},
  year = {2020},
  month = {Dec},
  publisher = {American Physical Society},
  doi = {10.1103/PhysRevLett.125.240502},
  url = {https://link.aps.org/doi/10.1103/PhysRevLett.125.240502}
}

@unpublished{Hill2021,
      title={Realization of arbitrary doubly-controlled quantum phase gates}, 
      author={Alexander D. Hill and Mark J. Hodson and Nicolas Didier and Matthew J. Reagor},
      year={2021},
      eprint={2108.01652},
      archivePrefix={arXiv},
      primaryClass={quant-ph},
}

@article{Khaneja2005,
    author = {Khaneja, Navin and Reiss, Timo and Kehlet, Cindie and Schulte-Herbr\"uggen, Thomas and Glaser, Steffen J.},
    title = "{Optimal control of coupled spin dynamics: design of NMR pulse sequences by gradient ascent algorithms}",
    doi = "10.1016/j.jmr.2004.11.004",
    journal = "J. Magn. Reson.",
    volume = "172",
    number = "2",
    pages = "296--305",
    year = "2005"
}

@unpublished{Klimov2020,
      title={The Snake Optimizer for Learning Quantum Processor Control Parameters}, 
      author={Paul V. Klimov and Julian Kelly and John M. Martinis and Hartmut Neven},
      year={2020},
      eprint={2006.04594},
      archivePrefix={arXiv},
      primaryClass={quant-ph}
}

@article{Jurcevic2021,
  title = {Demonstration of quantum volume 64 on a superconducting quantum computing system},
  volume = {6},
  ISSN = {2058-9565},
  url = {http://dx.doi.org/10.1088/2058-9565/abe519},
  DOI = {10.1088/2058-9565/abe519},
  number = {2},
  journal = {Quantum Sci. Technol.},
  publisher = {IOP Publishing},
  author = {Jurcevic,  Petar and others},
  year = {2021},
  month = mar,
  pages = {025020}
}

@online{Almeida2024,
  author       = {Daniella Garcia Almeida and Kaelyn Ferris and Naoki Kanazawa and Blake Johnson and Robert Davis},
  title        = {New fractional gates reduce circuit depth for utility-scale workloads},
  year         = {2024},
  month        = {November},
  day          = {7},
  howpublished  = {\url{https://www.ibm.com/quantum/blog/fractional-gates}},
  note         = {(accessed: 2025-11-12)},
  organization = {IBM Quantum}
}

@article{Alexander2020,
  title = {Qiskit pulse: programming quantum computers through the cloud with pulses},
  volume = {5},
  ISSN = {2058-9565},
  url = {http://dx.doi.org/10.1088/2058-9565/aba404},
  DOI = {10.1088/2058-9565/aba404},
  number = {4},
  journal = {Quantum Sci. Technol.},
  publisher = {IOP Publishing},
  author = {Alexander,  Thomas and Kanazawa,  Naoki and Egger,  Daniel J and Capelluto,  Lauren and Wood,  Christopher J and Javadi-Abhari,  Ali and C McKay,  David},
  year = {2020},
  month = aug,
  pages = {044006}
}

@article{Baum2021,
  title = {Experimental Deep Reinforcement Learning for Error-Robust Gate-Set Design on a Superconducting Quantum Computer},
  author = {Baum, Yuval and Amico, Mirko and Howell, Sean and Hush, Michael and Liuzzi, Maggie and Mundada, Pranav and Merkh, Thomas and Carvalho, Andre R.R. and Biercuk, Michael J.},
  journal = {PRX Quantum},
  volume = {2},
  issue = {4},
  pages = {040324},
  numpages = {12},
  year = {2021},
  month = {Nov},
  publisher = {American Physical Society},
  doi = {10.1103/PRXQuantum.2.040324},
  url = {https://link.aps.org/doi/10.1103/PRXQuantum.2.040324}
}

@article{Carvalho2021,
  title = {Error-Robust Quantum Logic Optimization Using a Cloud Quantum Computer Interface},
  author = {Carvalho, Andre R. R. and Ball, Harrison and Biercuk, Michael J. and Hush, Michael R. and Thomsen, Felix},
  journal = {Phys. Rev. Appl.},
  volume = {15},
  issue = {6},
  pages = {064054},
  numpages = {19},
  year = {2021},
  month = {Jun},
  publisher = {American Physical Society},
  doi = {10.1103/PhysRevApplied.15.064054},
  url = {https://link.aps.org/doi/10.1103/PhysRevApplied.15.064054}
}

@article{Machnes2018,
    author = "Machnes, Shai and Ass\'emat, Elie and Tannor, David and Wilhelm, Frank K.",
    title = "{Tunable, Flexible, and Efficient Optimization of Control Pulses for Practical Qubits}",
    doi = "10.1103/PhysRevLett.120.150401",
    journal = "Phys. Rev. Lett.",
    volume = "120",
    number = "15",
    pages = "150401",
    year = "2018"
}

@Article{Warren2023,
author={Warren, Christopher W.
and Fern{\'a}ndez-Pend{\'a}s, Jorge
and Ahmed, Shahnawaz
and Abad, Tahereh
and Bengtsson, Andreas
and Bizn{\'a}rov{\'a}, Janka
and Debnath, Kamanasish
and Gu, Xiu
and Kri{\v{z}}an, Christian
and Osman, Amr
and Fadavi Roudsari, Anita
and Delsing, Per
and Johansson, G{\"o}ran
and Frisk Kockum, Anton
and Tancredi, Giovanna
and Bylander, Jonas},
title={Extensive characterization and implementation of a family of three-qubit gates at the coherence limit},
journal={npj Quantum Information},
year={2023},
month={May},
day={04},
volume={9},
number={1},
pages={44},
issn={2056-6387},
doi={10.1038/s41534-023-00711-x},
url={https://doi.org/10.1038/s41534-023-00711-x}
}

@article{Chow2011,
  title = {Simple All-Microwave Entangling Gate for Fixed-Frequency Superconducting Qubits},
  author = {Chow, Jerry M. and C\'orcoles, A. D. and Gambetta, Jay M. and Rigetti, Chad and Johnson, B. R. and Smolin, John A. and Rozen, J. R. and Keefe, George A. and Rothwell, Mary B. and Ketchen, Mark B. and Steffen, M.},
  journal = {Phys. Rev. Lett.},
  volume = {107},
  issue = {8},
  pages = {080502},
  numpages = {5},
  year = {2011},
  month = {Aug},
  publisher = {American Physical Society},
  doi = {10.1103/PhysRevLett.107.080502},
  url = {https://link.aps.org/doi/10.1103/PhysRevLett.107.080502}
}

@article{Sheldon2016,
  title = {Procedure for systematically tuning up cross-talk in the cross-resonance gate},
  author = {Sheldon, Sarah and Magesan, Easwar and Chow, Jerry M. and Gambetta, Jay M.},
  journal = {Phys. Rev. A},
  volume = {93},
  issue = {6},
  pages = {060302},
  numpages = {5},
  year = {2016},
  month = {Jun},
  publisher = {American Physical Society},
  doi = {10.1103/PhysRevA.93.060302},
  url = {https://link.aps.org/doi/10.1103/PhysRevA.93.060302}
}

@article{Magesan2020,
  title = {Effective Hamiltonian models of the cross-resonance gate},
  author = {Magesan, Easwar and Gambetta, Jay M.},
  journal = {Phys. Rev. A},
  volume = {101},
  issue = {5},
  pages = {052308},
  numpages = {15},
  year = {2020},
  month = {May},
  publisher = {American Physical Society},
  doi = {10.1103/PhysRevA.101.052308},
  url = {https://link.aps.org/doi/10.1103/PhysRevA.101.052308}
}

@article{Sundaresan2020,
  title = {Reducing Unitary and Spectator Errors in Cross Resonance with Optimized Rotary Echoes},
  author = {Sundaresan, Neereja and Lauer, Isaac and Pritchett, Emily and Magesan, Easwar and Jurcevic, Petar and Gambetta, Jay M.},
  journal = {PRX Quantum},
  volume = {1},
  issue = {2},
  pages = {020318},
  numpages = {23},
  year = {2020},
  month = {Dec},
  publisher = {American Physical Society},
  doi = {10.1103/PRXQuantum.1.020318},
  url = {https://link.aps.org/doi/10.1103/PRXQuantum.1.020318}
}

@article{Stenger2021,
  title = {Simulating the dynamics of braiding of Majorana zero modes using an IBM quantum computer},
  author = {Stenger, John P. T. and Bronn, Nicholas T. and Egger, Daniel J. and Pekker, David},
  journal = {Phys. Rev. Res.},
  volume = {3},
  issue = {3},
  pages = {033171},
  numpages = {11},
  year = {2021},
  month = {Aug},
  publisher = {American Physical Society},
  doi = {10.1103/PhysRevResearch.3.033171},
  url = {https://link.aps.org/doi/10.1103/PhysRevResearch.3.033171}
}

@article{Earnest2021,
  title = {Pulse-efficient circuit transpilation for quantum applications on cross-resonance-based hardware},
  author = {Earnest, Nathan and Tornow, Caroline and Egger, Daniel J.},
  journal = {Phys. Rev. Res.},
  volume = {3},
  issue = {4},
  pages = {043088},
  numpages = {11},
  year = {2021},
  month = {Oct},
  publisher = {American Physical Society},
  doi = {10.1103/PhysRevResearch.3.043088},
  url = {https://link.aps.org/doi/10.1103/PhysRevResearch.3.043088}
}

@article{Peterson2022,
   title={Optimal synthesis into fixed XX interactions},
   volume={6},
   ISSN={2521-327X},
   url={http://dx.doi.org/10.22331/q-2022-04-27-696},
   DOI={10.22331/q-2022-04-27-696},
   journal={Quantum},
   publisher={Verein zur Forderung des Open Access Publizierens in den Quantenwissenschaften},
   author={Peterson, Eric C. and Bishop, Lev S. and Javadi-Abhari, Ali},
   year={2022},
   month=apr, pages={696} }

@unpublished{McKinney2025,
      title={Two-Qubit Gate Synthesis via Linear Programming for Heterogeneous Instruction Sets}, 
      author={Evan McKinney and Lev S. Bishop},
      year={2025},
      eprint={2505.00543},
      archivePrefix={arXiv},
      primaryClass={quant-ph}
}

@article{Shabani2011,
  title = {Efficient Measurement of Quantum Dynamics via Compressive Sensing},
  author = {Shabani, A. and Kosut, R. L. and Mohseni, M. and Rabitz, H. and Broome, M. A. and Almeida, M. P. and Fedrizzi, A. and White, A. G.},
  journal = {Phys. Rev. Lett.},
  volume = {106},
  issue = {10},
  pages = {100401},
  numpages = {4},
  year = {2011},
  month = {Mar},
  publisher = {American Physical Society},
  doi = {10.1103/PhysRevLett.106.100401},
  url = {https://link.aps.org/doi/10.1103/PhysRevLett.106.100401}
}

@article{Flammia2012,
  title = {Quantum tomography via compressed sensing: error bounds,  sample complexity and efficient estimators},
  volume = {14},
  ISSN = {1367-2630},
  url = {http://dx.doi.org/10.1088/1367-2630/14/9/095022},
  DOI = {10.1088/1367-2630/14/9/095022},
  number = {9},
  journal = {New J. Phys.},
  publisher = {IOP Publishing},
  author = {Flammia,  Steven T and Gross,  David and Liu,  Yi-Kai and Eisert,  Jens},
  year = {2012},
  month = sep,
  pages = {095022}
}

@article{Gutoski2014,
    author = {Gutoski, Gus and Johnston, Nathaniel},
    title = {Process tomography for unitary quantum channels},
    journal = {J. Math. Phys.},
    volume = {55},
    number = {3},
    pages = {032201},
    year = {2014},
    month = {03},
    issn = {0022-2488},
    doi = {10.1063/1.4867625},
    url = {https://doi.org/10.1063/1.4867625}
}

@article{Baldwin2014,
  title = {Quantum process tomography of unitary and near-unitary maps},
  author = {Baldwin, Charles H. and Kalev, Amir and Deutsch, Ivan H.},
  journal = {Phys. Rev. A},
  volume = {90},
  issue = {1},
  pages = {012110},
  numpages = {10},
  year = {2014},
  month = {Jul},
  publisher = {American Physical Society},
  doi = {10.1103/PhysRevA.90.012110},
  url = {https://link.aps.org/doi/10.1103/PhysRevA.90.012110}
}

@article{Nielsen2021,
   title={Gate Set Tomography},
   volume={5},
   ISSN={2521-327X},
   url={http://dx.doi.org/10.22331/q-2021-10-05-557},
   DOI={10.22331/q-2021-10-05-557},
   journal={Quantum},
   publisher={Verein zur Forderung des Open Access Publizierens in den Quantenwissenschaften},
   author={Nielsen, Erik and Gamble, John King and Rudinger, Kenneth and Scholten, Travis and Young, Kevin and Blume-Kohout, Robin},
   year={2021},
   month=oct, pages={557} }

@article{Ball2021,
  title = {Software tools for quantum control: improving quantum computer performance through noise and error suppression},
  volume = {6},
  ISSN = {2058-9565},
  url = {http://dx.doi.org/10.1088/2058-9565/abdca6},
  DOI = {10.1088/2058-9565/abdca6},
  number = {4},
  journal = {Quantum Sci. Technol.},
  publisher = {IOP Publishing},
  author = {Ball,  Harrison and Biercuk,  Michael J and Carvalho,  Andre R R and Chen,  Jiayin and Hush,  Michael and De Castro,  Leonardo A and Li,  Li and Liebermann,  Per J and Slatyer,  Harry J and Edmunds,  Claire and Frey,  Virginia and Hempel,  Cornelius and Milne,  Alistair},
  year = {2021},
  month = sep,
  pages = {044011}
}

@article{Wei2024,
  title = {Characterizing non-Markovian off-resonant errors in quantum gates},
  author = {Wei, Ken Xuan and Pritchett, Emily and Zajac, David M. and McKay, David C. and Merkel, Seth},
  journal = {Phys. Rev. Appl.},
  volume = {21},
  issue = {2},
  pages = {024018},
  numpages = {19},
  year = {2024},
  month = {Feb},
  publisher = {American Physical Society},
  doi = {10.1103/PhysRevApplied.21.024018},
  url = {https://link.aps.org/doi/10.1103/PhysRevApplied.21.024018}
}

@Article{Gross2024,
author={Gross, Jonathan A.
and Genois, {\'E}lie
and Debroy, Dripto M.
and Zhang, Yaxing
and Mruczkiewicz, Wojciech
and Cian, Ze-Pei
and Jiang, Zhang},
title={Characterizing coherent errors using matrix-element amplification},
journal={npj Quantum Inf.},
year={2024},
month={Nov},
day={22},
volume={10},
number={1},
pages={123},
issn={2056-6387},
doi={10.1038/s41534-024-00917-7},
url={https://doi.org/10.1038/s41534-024-00917-7}
}

@article{Makhlin2002, volume={1},
   ISSN={1570-0755},
   url={http://dx.doi.org/10.1023/A:1022144002391},
   DOI={10.1023/a:1022144002391},
   number={4},
   journal={Quantum Inf. Process.},
   publisher={Springer Science and Business Media LLC},
   author={Makhlin, Yuriy},
   year={2002},
   pages={243–252} }

@article{Zhang2003,
  title = {Geometric theory of nonlocal two-qubit operations},
  author = {Zhang, Jun and Vala, Jiri and Sastry, Shankar and Whaley, K. Birgitta},
  journal = {Phys. Rev. A},
  volume = {67},
  issue = {4},
  pages = {042313},
  numpages = {18},
  year = {2003},
  month = {Apr},
  publisher = {American Physical Society},
  doi = {10.1103/PhysRevA.67.042313},
  url = {https://link.aps.org/doi/10.1103/PhysRevA.67.042313}
}

@ARTICLE{Lubinski2023,
  author={Lubinski, Thomas and Johri, Sonika and Varosy, Paul and Coleman, Jeremiah and Zhao, Luning and Necaise, Jason and Baldwin, Charles H. and Mayer, Karl and Proctor, Timothy},
  journal={IEEE Trans. Quantum Eng.}, 
  title={Application-Oriented Performance Benchmarks for Quantum Computing}, 
  year={2023},
  volume={4},
  number={},
  pages={1-32},
  doi={10.1109/TQE.2023.3253761}}

@article{Sawaya2024,
   title={HamLib: A library of Hamiltonians for benchmarking quantum algorithms and hardware},
   volume={8},
   ISSN={2521-327X},
   url={http://dx.doi.org/10.22331/q-2024-12-11-1559},
   DOI={10.22331/q-2024-12-11-1559},
   journal={Quantum},
   publisher={Verein zur Forderung des Open Access Publizierens in den Quantenwissenschaften},
   author={Sawaya, Nicolas PD and others},
   year={2024},
   month=dec, pages={1559} }

@article{Lao2022,
  title = {Software mitigation of coherent two-qubit gate errors},
  volume = {7},
  ISSN = {2058-9565},
  url = {http://dx.doi.org/10.1088/2058-9565/ac57f1},
  DOI = {10.1088/2058-9565/ac57f1},
  number = {2},
  journal = {Quantum Sci. Technol.},
  publisher = {IOP Publishing},
  author = {Lao,  Lingling and Korotkov,  Alexander and Jiang,  Zhang and Mruczkiewicz,  Wojciech and O’Brien,  Thomas E and Browne,  Dan E},
  year = {2022},
  month = mar,
  pages = {025021}
}

@article{Rezakhani2004,
  title = {Characterization of two-qubit perfect entanglers},
  author = {Rezakhani, A. T.},
  journal = {Phys. Rev. A},
  volume = {70},
  issue = {5},
  pages = {052313},
  numpages = {9},
  year = {2004},
  month = {Nov},
  publisher = {American Physical Society},
  doi = {10.1103/PhysRevA.70.052313},
  url = {https://link.aps.org/doi/10.1103/PhysRevA.70.052313}
}

@INPROCEEDINGS {Lin2022,
author = { Lin, Sophia Fuhui and Sussman, Sara and Duckering, Casey and Mundada, Pranav S. and Baker, Jonathan M. and Kumar, Rohan S. and Houck, Andrew A. and Chong, Frederic T. },
booktitle = { 2022 55th IEEE/ACM International Symposium on Microarchitecture},
title = {{ Let Each Quantum Bit Choose Its Basis Gates }},
year = {2022},
volume = {},
ISSN = {},
pages = {1042-1058},
doi = {10.1109/MICRO56248.2022.00075},
url = {https://doi.ieeecomputersociety.org/10.1109/MICRO56248.2022.00075},
publisher = {IEEE Computer Society},
address = {Los Alamitos, CA, USA},
month =Oct}

@unpublished{Tucci2005,
      title={An Introduction to {Cartan's} {KAK} Decomposition for {QC} Programmers}, 
      author={Robert R. Tucci},
      year={2005},
      eprint={quant-ph/0507171},
      archivePrefix={arXiv},
      primaryClass={quant-ph}
}

@article{Zhang2004,
  title = {Minimum Construction of Two-Qubit Quantum Operations},
  author = {Zhang, Jun and Vala, Jiri and Sastry, Shankar and Whaley, K. Birgitta},
  journal = {Phys. Rev. Lett.},
  volume = {93},
  issue = {2},
  pages = {020502},
  numpages = {4},
  year = {2004},
  month = {Jul},
  publisher = {American Physical Society},
  doi = {10.1103/PhysRevLett.93.020502},
  url = {https://link.aps.org/doi/10.1103/PhysRevLett.93.020502}
}

@unpublished{Sugawara2025,
      title={SU(4) gate design via unitary process tomography: its application to cross-resonance based superconducting quantum devices}, 
      author={Michihiko Sugawara and Takahiko Satoh},
      year={2025},
      eprint={2503.09343},
      archivePrefix={arXiv},
      primaryClass={quant-ph}
}

@article{Zhang2005,
  title = {Conditions for optimal construction of two-qubit nonlocal gates},
  author = {Zhang, Yong-Sheng and Ye, Ming-Yong and Guo, Guang-Can},
  journal = {Phys. Rev. A},
  volume = {71},
  issue = {6},
  pages = {062331},
  numpages = {6},
  year = {2005},
  month = {Jun},
  publisher = {American Physical Society},
  doi = {10.1103/PhysRevA.71.062331},
  url = {https://link.aps.org/doi/10.1103/PhysRevA.71.062331}
}

\clearpage

\appendix

\section{Computing and optimizing Makhlin local invariants}\label{app:makhlin-invariants}

In \cref{sec: Compiler architecture design for efficient gates}, our general synthesis technique hinges on calculating and optimizing the difference between the Makhlin invariants of two unitary matrices. Here, we explicitly define these invariants, and justify their convenience for gradient-based numerical optimization.

Theorem 2 from~\cite{Makhlin2002} provides a procedure to get the complete set of local invariants of a two-qubit gate $U$. To calculate these invariants, one needs the magic matrix $Q$ which transforms computational basis states into the Bell basis,
\begin{align} \label{magic matrix Q defined}
        Q = \begin{pmatrix}
            1 & 0 & 0 & i\\
            0 & i & 1 & 0\\
            0 & i & -1 & 0\\
            1 & 0 & 0 & -i\\
        \end{pmatrix} \, .
\end{align}
In this basis, the unitary is first transformed into $ U_B=Q^\dag U Q $, and then the matrix $m=U^T_B U_B$ is defined. Then the invariants are obtained as
\begin{align}  \label{local invariants stated}
    g_1( U ) = \ \frac{1}{4} \ \text{tr} (m), \ g_2( U ) = \ \ \frac{1}{4} \left(\text{tr}^2(m) - \text{tr}(m^2) \right) \, .
\end{align}
The first invariant is complex-valued, and the second real-valued. The Makhlin invariants of the canonical gate $\can(c_1,c_2,c_3)$ can be calculated as
\begin{align}  \label{local invariants stated for canonical gate}
    g_1(\can(c_1,c_2,c_3)) &= \cos c_1 \cos c_2 \cos c_3 \\
    &\hspace{1em}-i \sin c_1 \sin c_2 \sin c_3 \, ,\\
    g_2(\can(c_1,c_2,c_3)) &= \cos 2 c_1 + \cos 2  c_2 + \cos 2  c_3 \, .
\end{align}

Optimizing the squared difference of the Makhlin invariants is amenable to gradient-based optimization methods because each invariant component of $\mathbf{g}(U)$ is simply a low-degree polynomial in the real and imaginary entries of the input matrix $U$. Computing $U_B$ is a fixed linear change of basis, $m$ is formed from a matrix multiplication, and the traces are sums of products of matrix elements, leading to a low-degree polynomial transformation. There are no branch cuts or non-differentiable points, with the gradient straightforwardly obtained in closed form or via automatic differentiation.

\section{Circuit synthesis using single-axis two-qubit gates}\label{app: single parameter identities}

We look in more detail at the key motif for synthesizing $\can(c_1, c_2, 0)$ using $\can(c_\text{eff}^{(k_1)}, 0, 0)$ and $\can(c_\text{eff}^{(k_2)}, 0, 0)$ for some characterized pulse indices $\mathbf{k}=(k_1, k_2)$. Recall we are considering a structure
\begin{align}
    \can(c_1, c_2, 0) =\; & (Z(\theta_1)  Z(\theta_2)) \cdot \can(c_\text{eff}^{(k_1)}, 0, 0) \\
    \cdot \, & (Z(\phi_1)  Z(\phi_2))\cdot\can(c_\text{eff}^{(k_2)}, 0, 0) \\
    \cdot \, &(Z(\theta_3)  Z(\theta_4)) \, .
\end{align}
For notational convenience we define $\Sigma_\text{eff} = c_\text{eff}^{(k_1)} + c_\text{eff}^{(k_2)}$, $\Delta_\text{eff} = c_\text{eff}^{(k_1)} - c_\text{eff}^{(k_2)}$, $\Sigma_\text{block} = c_1 + c_2$, and $\Delta_\text{block} = c_1 - c_2$.
By algebraically matching Makhlin invariants as per \cref{eq:makhlin-equivalence-single-axis-pulse}, we find that $\phi_1 = \phi_\Delta + \phi_\Sigma$ and $\phi_2 = \phi_\Delta - \phi_\Sigma$, where
\begin{align*}
    \cos 2 \phi_a = \frac{\cos \Sigma_\text{eff} + \cos \Delta_\text{eff} - 2 \cos a}{\cos \Delta_\text{eff} - \cos \Sigma_\text{eff}} \, .
\end{align*}
The two $\phi_a$ are real-valued when the RHS are between -1 and +1, leading to the constraints $\Sigma_\text{eff} \geq \Sigma_\text{block}$ and $\Delta_\text{eff} \leq \Delta_\text{block}$. Finally, solving for the outer rotation angles by matching matrix elements gives
\begin{align}
    \theta_1 &= \gamma_{\Delta, 1} + \gamma_{\Delta, 2} + \gamma_{\Sigma, 1} + \gamma_{\Sigma, 2} \, , \\
    \theta_2 &= \gamma_{\Delta, 1} + \gamma_{\Delta, 2} - \gamma_{\Sigma, 1} - \gamma_{\Sigma, 2} \, , \\
    \theta_3 &= \gamma_{\Delta, 1} - \gamma_{\Delta, 2} + \gamma_{\Sigma, 1} - \gamma_{\Sigma, 2} \, ,\\
    \theta_4 &= \gamma_{\Delta, 1} - \gamma_{\Delta, 2} - \gamma_{\Sigma, 1} + \gamma_{\Sigma, 2} \, ,
\end{align}
where $\tan 2 \gamma_{a, i} = m_i \tan \phi_a$ with $m_1 = -\cos (\Delta_\text{eff}/2) / \cos (\Sigma_\text{eff} / 2)$ and $m_2 = {\sin(\Delta_\text{eff} / 2)}/{\sin(\Sigma_\text{eff} / 2)}$.
Thus, we have synthesized the canonical gate $\can(c_1, c_2, 0)$ using two single-axis canonical gates and single-qubit $Z$ rotations.

The other non-trivial case to consider is that of a general target canonical gate $\can(c_1, c_2, c_3)$ using three single-axis pulses. As explained in the main text, this can be achieved using two applications of the above motif, by first synthesizing $\can(c_1, c_2, 0)$ into $\can(c_\text{eff}^{(k_1)}, 0, 0)$ and $\can(\delta, 0, 0)$, and then after rearranging, synthesizing $\can(\delta, c_3, 0)$ into $\can(c_\text{eff}^{(k_2)}, 0, 0)$ and $\can(c_\text{eff}^{(k_3)}, 0, 0)$. We now prove the set of conditions given in the main text for a valid $\delta$ to exist, and thus for the target to be synthesizable using a given $\mathbf{k}$.
\begin{proof}
    From the two applications of the two-pulse motif, we have
    \begin{align}
        c_\text{eff}^{(k_1)} + \delta &\geq c_1 + c_2 \, , \\
        c_\text{eff}^{(k_1)} - \delta &\leq c_1 - c_2 \, , \\
        c_\text{eff}^{(k_2)} + c_\text{eff}^{(k_3)} &\geq \delta + c_3 \, , \\
        c_\text{eff}^{(k_2)} - c_\text{eff}^{(k_3)} &\leq \delta - c_3 \, .
    \end{align}
    Rearranging the four inequalities to isolate $\delta$ in each case, we must have $L \leq \delta \leq U$, where
    \begin{align*}
        L &= \max\{c_1 + c_2 - c_\text{eff}^{(k_1)}, c_\text{eff}^{(k_1)} - c_1 + c_2 , \\ &\qquad\qquad c_\text{eff}^{(k_2)} - c_\text{eff}^{(k_3)} + c_3 \} \, , \\
        U &= c_\text{eff}^{(k_2)} + c_\text{eff}^{(k_3)} - c_3 \,.
    \end{align*}
    Since every element in $L$ must not exceed $U$ for the valid $\delta$ region to be non-empty, we obtain the equivalent set of conditions
    \begin{align}
        c_1 + c_2 + c_3 &\leq c_\text{eff}^{(k_1)} +  c_\text{eff}^{(k_2)} + c_\text{eff}^{(k_3)}\, , \\
         - c_1 + c_2 + c_3 &\leq - c_\text{eff}^{(k_1)} + c_\text{eff}^{(k_2)} + c_\text{eff}^{(k_3)} \, ,\\
         c_3 &\leq c_\text{eff}^{(k_3)} \, ,
    \end{align}
    as required.
\end{proof}
In practice when performing synthesis, the value of $\delta$ can be fixed to (for example) $\delta=U$.

\section{Controlled pulses}
\label{sec:app:ceff-derivation}

Here we briefly verify that controlled pulses are single-axis pulses, and give the canonical coordinates $(c_1, 0, 0)$.

First, see that from the relation between Makhlin invariants and canonical coordinates given in \cref{app:makhlin-invariants}, $c_2$ and $c_3$ are zero if and only if $\Im{g_1} = 0$ and $g_2 - 2\Re{g_1} = 1$. This can be explicitly calculated and found to be the case for the block-diagonal unitary matrix of an arbitrary controlled pulse. We find that
\begin{equation}
    g_1 = \cos u \cos v + \frac{\mathbf{u} \cdot \mathbf{v}}{u \, v} \sin u \sin v = \cos c_1
\end{equation}
where $u = \norm{\mathbf{u}}$ and $v=\norm{\mathbf{v}}$. This produces the canonical coordinate $c_1$. Curiously, this expression coincides with the spherical law of cosines, where $c_1$ is the great-circle distance (central angle) between the unit vectors $\hat{\mathbf{u}}$ and $\hat{\mathbf{v}}$ on the unit sphere.

\end{document}